\newif\ifanon
\anonfalse

\newif\ifonecolumn
\onecolumnfalse

\newif\ifdarkdoc
\darkdocfalse

\newcommand{\inOneTwoColumn}[2]{\ifonecolumn #1 \else #2 \fi}  

\inOneTwoColumn{
\newcommand{\SETUPcolumn}{onecolumn}
}{
\newcommand{\SETUPcolumn}{twocolumn}
}

\documentclass[
superscriptaddress,
nofootinbib,
amsmath,amssymb,
aps,
pre,
\SETUPcolumn,
]{revtex4-2}

\usepackage{xcolor,colortbl}
\ifdarkdoc
\pagecolor[rgb]{0.1,0.1,0.1}
\color[rgb]{1,1,1}
\fi

\usepackage{subfiles}
\usepackage{etoolbox}

\usepackage{graphicx}
\graphicspath{{Figures/serif/}}

\usepackage{dcolumn}
\usepackage{etoolbox,siunitx}
\robustify\bfseries
\usepackage{bm}
\usepackage{hyperref}
\usepackage{xfp} 
\usepackage{booktabs}

\usepackage[capitalize]{cleveref} 

\hypersetup{colorlinks=true, 
linkcolor=blue,
filecolor=blue,      
urlcolor=blue,
citecolor=blue,
}

\newcolumntype{C}[1]{>{\centering\arraybackslash}p{#1}}

\usepackage{upgreek}
\usepackage{bm}
\newcommand{\vect}[1]{\boldsymbol{\mathbf{#1}}}

\usepackage{Styles/2021_MFT_macros}
\usepackage{Styles/EN_GB}

\newcommand{\mdUoBathAdd}{University of Bath, Bath BA2 7AY, United Kingdom}

\newcommand{\docPreAbstract}{
\preprint{APS/123-QED}

\title{2T-POT Hawkes model for left- and right-tail conditional quantile forecasts of financial log-returns: out-of-sample comparison of conditional EVT models}

\ifanon
\else
\author{Matthew F. Tomlinson}
	\email{mft28@bath.ac.uk}
\affiliation{Department of Physics, \mdUoBathAdd}
\affiliation{Centre for Networks and Collective Behaviour, \mdUoBathAdd}

\author{David Greenwood}
\affiliation{
CheckRisk LLP, 4 Miles's Buildings, George Street, Bath BA1 2QS, United Kingdom}

\author{Marcin Mucha-Kruczy\'{n}ski}%
\affiliation{Department of Physics, \mdUoBathAdd}
\affiliation{Centre for Nanoscience and Nanotechnology, \mdUoBathAdd}
\fi

\iffalse
\date{\today}
\else
\date{14th October 2022}
\fi
}

\newcommand{\docPostAbstract}{
\keywords{Hawkes processes, GARCH-EVT, conditional extreme value theory, value-at-risk, expected shortfall, leverage effect}
\maketitle
}

\newcommand{\FTABdatastat}{
\begin{ruledtabular}
\footnotesize
\begin{tabular}{cccccccccccccc}
\TABdatastat
\end{tabular}
\end{ruledtabular}
}

\newcommand{\FTABrCTagg}{
\begin{ruledtabular}
\footnotesize
\begin{tabular}{cccrrrrrrrrrrrrr}
\TABrCTagg
\end{tabular}
\end{ruledtabular}
}
\newcommand{\FTABrCTevt}{
\begin{ruledtabular}
\footnotesize
\begin{tabular}{cccrrrrrrrrrrr}
\TABrCTevt
\end{tabular}
\end{ruledtabular}
}

\newcommand{\FTABpLRci}{
\begin{ruledtabular}
\footnotesize
\begin{tabular}{cccccccccccccc}
\TABpLRci
\end{tabular}
\end{ruledtabular}
}
\newcommand{\FTABpLRc}{
\begin{ruledtabular}
\footnotesize
\begin{tabular}{cccccccccccccc}
\TABpLRc
\end{tabular}
\end{ruledtabular}
}
\newcommand{\FTABpLRt}{
\begin{ruledtabular}
\footnotesize
\begin{tabular}{cccccccccccccc}
\TABpLRt
\end{tabular}
\end{ruledtabular}
}

\newcommand{\docAcknowledgeOpen}{
\begin{acknowledgements}
}
\newcommand{\docAcknowledgeClose}{
\end{acknowledgements}
}

\ifanon
\newcommand{\BIBprint}{
\bibliography{Bibliography/references_abv_anon}
}
\else
\newcommand{\BIBprint}{
\bibliography{Bibliography/references_abv}
}
\fi

\begin{document}

\docPreAbstract
\begin{abstract}
Conditional extreme value theory (EVT) methods promise enhanced forecasting of the extreme tail events that often dominate systemic risk. We present an improved two-tailed peaks-over-threshold (2T-POT) Hawkes model that is adapted for conditional quantile forecasting in both the left and right tails of a univariate time series. This is applied to the daily log-returns of six large cap indices. We also take the unique step of fitting the model at multiple exceedance thresholds (from the 1.25\% to 25.00\% mirrored quantiles). Quantitatively similar asymmetries in Hawkes parameters are found across all six indices, adding further empirical support to a temporal leverage effect in financial price time series in which the impact of losses is not only larger but also more immediate. Out-of-sample backtests find that our 2T-POT Hawkes model is more reliably accurate than the GARCH-EVT model when forecasting (mirrored) value-at-risk and expected shortfall at the 5\% coverage level and below. This suggests that asymmetric Hawkes-type arrival dynamics are a better approximation of the true data generating process for extreme daily log-returns than GARCH-type conditional volatility; our 2T-POT Hawkes model therefore presents a better performing alternative for financial risk mode\ENll{}ing.
\end{abstract}
\docPostAbstract

\section{\label{sec:intro}Introduction}

A notable feature of many complex systems is that outcomes are often influenced more by rare extreme events than by more typical fluctuations \citep{Sornette2006}. As a result, these extreme tail events often dominate the associated systemic risk, which makes accurate forecasting of them a vital objective in many disciplines. Extreme value theory (EVT) presents an approach to this problem in which asymptotic tail behaviour is mode\ENll{}ed independently from the typical ``bulk'' fluctuations, under the justification that the two are often generated by distinct mechanisms \citep{Coles2001, DeHaan2006}. In practical terms, this means that values beyond a defined threshold are classified as extremes and are described by an exceedance distribution that is fitted independently from the sub\ENhyph{}threshold (i.e.\ENLA{} non\ENhyphNon{}extreme or \textit{bulk}) distribution. The simplest application is made when the data generating process for extreme events is stationary. In this case, not only is the distribution of exceedances itself unconditional, but the intensity (i.e.\ENLA{} arrival rate in time) of events from this distribution is constant. Together, these two conditions mean that arrivals of extreme events beyond any arbitrary threshold within the tail distribution occur in time according to a homogeneous Poisson point process (as defined by a constant intensity). However, in some systems of interest such as financial markets, complex dynamics and feedbacks give rise to non\ENhyphNon{}stationary behavi\ENou{}r that means the conditions of constant intensity and of an unconditional distribution of exceedance magnitudes cannot be assumed to hold. In such cases, extreme events tend to cluster together in time and their arrival intensity is often correlated with their magnitude; descriptions of these systems demand the development of conditional EVT methods that can account for these effects. 

This raises a question about whether these dynamics should also be treated independently from the bulk behaviour or whether they should be related to the conditional moments of the full distribution. In the context of finance, both of these approaches have been used to describe extreme price changes, as measured by extreme log-returns. A seminal example of the latter approach is the GARCH-EVT model created by \cite{McNeil2000}. This is an extension to the family of generali\ENz{}ed autoregressive conditional heteroscedasticity (GARCH) processes -- a ubiquitous class of reduced form models in financial analysis in which log-returns are described by independent and identically distributed (i.i.d.\ENLA{}) innovations (i.e.\ENLA{} white noise) scaled by the conditional heteroscedasticity, more often called the \textit{volatility} \citep{Ruppert2015}. The GARCH-EVT model appends a generali\ENz{}ed Pareto (GP) distribution to the innovation distribution of the GARCH process, in order to account for the residual heavy tails that are typically observed in log-returns when fitted with a standard (i.e.\ENLA{} non-EVT) GARCH process \citep{Cont2001}. It follows that the intensity and magnitude of extreme events are then simply functions of the conditional volatility. Conversely, a pure conditional EVT model in which the dynamics of threshold-exceeding extremes are purely self-contained is presented by the peaks-over-threshold (POT) Hawkes model. This combines an exceedance distribution with a self-exciting Hawkes point process to describe the time inhomogeneous arrivals of tail events. The Hawkes point process, which first emerged as a stochastic model for the self-reflexive pattern of foreshocks and aftershocks that decorate major seismic activity \citep{Hawkes1971b, Hawkes1971a, Adamopoulos1976, Shcherbakov2019}, is defined by past events causing a time-decaying increase in the intensity of future events \citep{Reinhart2018, Hawkes2018}. This approach has found broad application to many systems characteri\ENz{}ed by activity bursts, including neural networks \citep{Pernice2012, Tannenbaum2017}, crime \citep{Mohler2014, Mohler2018}, conflict \citep{Short2014, Johnson2018}, epidemics \citep{Chiang2022}, social media \citep{Fujita2018}, and financial markets \citep{Hawkes2018, Hawkes2020, Bacry2012, Bacry2015, Filimonov2012, Hardiman2013, Hardiman2014}. 

\cite{Chavez-Demoulin2005} were the first to apply a POT Hawkes model to extreme log-returns of a financial price time series, which they defined as those less than the 10\% in-sample quantile. In their approach, a Hawkes point process describes the arrival intensity of threshold exceeding log-returns, while the excess magnitudes of these events are described by an unconditional GP distribution. Subsequent literature has developed this work, by proposing alternative parametri\ENz{}ations \citep{Gresnigt2015, BienBarkowska2020}, incorporating exogenous signals \citep{Rambaldi2015} and through multivariate extensions that describe contagious shocks between different price series \citep{Embrechts2011, Grothe2014, Ait-Sahalia2015}. In this literature, the POT Hawkes model has almost always been used to describe the left-tail (i.e.\ENLA{} extreme losses) exclusively, but this neglects that log-returns also experience right-tail extremes (extreme gains)\footnote{In the few cases where right-tail extremes have been considered in the POT Hawkes literature, assumptions of left-right symmetry have been made. \cite{Embrechts2011} treat left- and right-tail extreme log-returns as distinct point processes in a bivariate Hawkes model, but they assume that the excitement produced by past events in either tail decays at the same rate (i.e.\ENLA{} they assume symmetry in the decay vector $\vect{\upbeta}$, such that $\beta^{\rsTL} = \beta^{\rsTR}$). \cite{Gresnigt2015} apply their univariate Hawkes model to threshold exceedances in the series of absolute log-returns $|X_t|$; this approach assumes that the two tails are symmetric in all respects.}. This reflects a similarly exclusive focus on left-tail risk in the broader financial risk literature. Indeed, the two most commonly used tail risk measures -- value-at-risk (VaR), which is a conditional quantile for the return of an investment over a given holding period, and the expected shortfall (ES), which is the expected value of the return given that it is less than the VaR -- are named and defined for the left-tail only. However, this focus disguises the importance of right-tail extremes, which are highly correlated with their left-tail counterparts \citep{Embrechts2011}, can either quickly mitigate the impact of their mirror opposites or present an equivalent risk (or opportunity) under certain investment strategies. This is especially apparent in the wake of the Covid-19 pandemic and the impact its outbreak had on global financial markets. On 2020-03-12, the S\&P 500 index suffered its worst daily loss since the infamous 1987 Black Monday crash, declining by $-9.5\%$. Five of the next twelve trading days saw losses of $-12.0\%$, $-5.2\%$, $-4.3\%$, $-2.9\%$, and $-3.4\%$. Crucially, however, the same period also saw daily gains of $+9.3\%$, $+6.0\%$, $+9.4\%$, $+6.2\%$, and $+3.4\%$. Without these strong upswings, the index would have lost a staggering $31.1\%$ of its value over this period rather than the $4.2\%$ aggregate drop that it did experience.

\cite{Tomlinson2021} developed the two-tail peaks-over-threshold (2T-POT) Hawkes model to investigate the interaction of and asymmetries between left- and right-tail extremes in financial log-returns. In this model, left- and right-tail threshold exceedances are described by an asymmetric self- and cross-exciting Hawkes-type arrivals process combined with asymmetric conditional GP tails. This can account for the time clustering of threshold exceeding extremes from both tails, the correlation between the arrival intensity and the magnitude of exceedances, the heavy tailed distributions of those excess magnitudes, and the propensity for all these features to exhibit left-versus-right tail asymmetry -- all of which are observed in the fluctuations of financial asset prices as measured by daily log-returns \citep{Cont2001, Davies2015, Ruppert2015, Tsay2010, Chicheportiche2014}. \cite{Tomlinson2021} applied their model to the daily log-returns of the S\&P 500 (SPX), with left- and right-tail extremes defined by thresholds set at the 2.5\% and 97.5\% quantiles, respectively. They found that extreme daily losses and gains shared a common conditional intensity: losses were found to contribute 2.2 times as much to this intensity overall and this contribution was found to decay 4.6 times as quickly. These two asymmetries are connected to the leverage effect: a well-known styli\ENz{}ed fact of financial markets, which states that volatility is negatively correlated with the sign of past log-returns \citep{Cont2001}. However, while the first of these asymmetries can be explained by conventional models of this effect, the latter cannot\footnote{Specifically, when the 2T-POT Hawkes model was fitted to simulated data generated by a GJR-GARCH process (defined in \cref{sssec:GARCH}) the impact asymmetry found in the branching vector $\vect{\upgamma}^{\rsTA}$ was reproduced but the time asymmetry found in the decay vector $\vect{\upbeta}$ was not \citep{Tomlinson2021}.}. If this temporal aspect of the leverage effect is demonstrated to be a common property of this and other classes of financial data, it would prove a novel insight into of one of the lowest-order non\ENhyphNon{}zero correlations in the price signal \citep{Chen2013}. This could provide an enhanced understanding of the generating mechanism for extreme log-returns from both tails, which, if exploited, could then improve the forecasting of these most consequential events.

In this paper, we directly compare the 2T-POT model with a set of GARCH models including GARCH-EVT, by testing accuracy of one step ahead forecasts of conditional quantile-based risk measures for all the approaches. To achieve this, we first build on the work of \cite{Tomlinson2021} and reparametri\ENz{}e the 2T-POT Hawkes model so that the expected average intensity replaces the exogenous background intensity as a fitting parameter. This halves the optimi\ENz{}ation time and enables a constraint that achieves a dimension reduction of \ENone{} at a negligible cost to the goodness of fit. The resulting improvements to the speed and reliability of the optimization procedure enable us to fit the 2T-POT Hawkes model to the daily log-returns of six international large cap equity indices over the in-sample period, \rsDateTrainStart{} to \rsDateTrainEnd{}, and, in a novel step in the POT Hawkes literature, repeat this over a wide range of exceedance thresholds, from the 1.25\% to 25.00\% mirrored in-sample quantiles. This allows us to determine the sensitivity of our results on threshold selection. Comparable asymmetries to those reported in \cite{Tomlinson2021} are found in the fitted parameters of all six indices across a wide range of thresholds; this provides evidence that the temporal leverage effect is a universal property of this class of assets.

We further expand the 2T-POT model by extending its support to the full distribution of log-returns via a subordinate bulk (i.e.\ENLA{} intra\ENhyph{}threshold) distribution that is conditional upon the Hawkes exceedance process. This guarantees that forecasts of conditional quantile-based risk measures are always defined at all coverage levels. We compute next step ahead VaR and ES forecasts for both the left and right tails and assess their accuracy and serial independence in the out-of-sample period, \rsDateTrainEnd{} to \rsDateAllEnd{}, through backtesting methods. This greatly expands upon similar analysis in previous literature \citep{BienBarkowska2020, Taylor2020, Echaust2020, Jalal2008}: both by extending the analysis to the right-tail and by evaluating forecasts over a much wider and finer range of coverage levels, from 0.25\% to 15.00\%. We find that our asymmetric 2T-POT Hawkes model produces the most reliably accurate forecasts in both the far left-tail (5\% quantile or less) and the far right-tail (95\% quantile or greater). This is not only of practical use, but also suggests that the asymmetric Hawkes-type arrival dynamics are a better approximation of the true data generating process for extreme daily log-returns than GARCH-type variance dynamics.

We present the reparametri\ENz{}ed 2T-POT model with the subordinate bulk distribution in \cref{sec:Model}. In this section, we also describe the GARCH-EVT model. The improved 2T-POT Hawkes model is fitted to the in-sample data in \cref{sec:Data}, where we then examine the estimated parameters across different thresholds. In \cref{sec:CT_QE}, we validate and compare the accuracy and serial independence of forecasts produced by the different models using backtesting methods in the out-of-sample data. In \cref{sec:reparam}, we derive the reparametri\ENz{}ation of the 2T-POT Hawkes model in terms of the expected intensity. In turn, \cref{sec:model_select} defines the likelihood function and optimi\ENz{}ation procedure for the model and shows the results of the likelihood ratio tests used for model selection.

\section{\label{sec:Model}Model specification}

\subsection{\label{ssec:QE}Conditional quantile-based risk measures}

We first explicitly define the conditional quantile-based risk measures that are subject to validation in \cref{sec:CT_QE} so that they can be explicitly defined for each model specified later in this section. When applied to the left-tail, these measures are known in the financial risk literature as value-at-risk and expected shortfall. However, because we generali\ENz{}e these to include equivalent measures for the right-tail, we adopt the more generic names: \textit{conditional quantile} and \textit{conditional violation expectation}.

\subsubsection{\label{sssec:Q}Conditional quantile}

If a discrete stochastic time series $X_t$ is generated according to the conditional cumulative distribution function (cdf) $F_{X,t}$, then the left-tail conditional quantile (i.e.\ENLA{} the value-at-risk) at the coverage level $a_q$ over the holding period from $t-1$ to $t$ is 
\begin{equation}
\label{eqn:VaR_L}
\rsVaR_{a_q, t}^{\rsTL} = F_{X,t}^{-1}{\left(a_q\right)} ,
\end{equation}
such that a fraction $a_q$ of $X_t$ are less than $\rsVaR_{a_q, t}^{\rsTL}$. Violations of the left-tail conditional quantile are indicated by the left-tail violation series
\begin{equation}
\label{eqn:VaR_I_L}
\rsVaRI_{a_q, t}^{\rsTL} = \mathcal{I}{\left[- \left(X_t - \rsVaR_{a_q, t}^{\rsTL}\right)\right]},
\end{equation}
where 
\begin{equation}
\label{eqn:Heaviside}
\mathcal{I}{\left[x\right]} = 
\begin{cases}
1, & x > 0 ,\\
0, & x \leq 0 ,\\
\end{cases} 
\end{equation}
is the Heaviside step function. The right-tail conditional quantile is similarly defined
\begin{equation}
\label{eqn:VaR_R}
\rsVaR_{a_q, t}^{\rsTR} = F_{X,t}^{-1}{\left(1 - a_q \right)} ,
\end{equation}
such that a fraction $a_q$ of $X_t$ are right-tail violations, as identified by the series, $\rsVaRI_{a_q, t}^{\rsTR} = \mathcal{I}{[+ (X_t - \rsVaR_{a_q, t}^{\rsTR})]}$.

Note that the superscripts $\rsTL$ and $\rsTR$ are used to denote the left- and right-tail, respectively. Henceforth the superscript $\rsTO$ is used to represent either tail (i.e.\ENLA{} either $\rsTL$ or $\rsTR$) in generic expressions and the superscript $\rsTA$ denotes a union of the left and right tails.

\subsubsection{\label{sssec:E}Conditional violation expectation}

The conditional quantile has the limitation of providing no information on the distribution beyond itself. In the context of finance, this has drawn the interest of many practitioners to expected shortfall (known as left-tail conditional violation expectation under our nomenclature) as an alternative risk measure. This has also been recogni\ENz{}ed at the level of regulation, where expected shortfall is now recommended as a risk measure by the Basel Committee on Banking Supervision \citep{Basel2016}. The left- and right-tail conditional violation expectations are defined as the conditional expectation of $X_t$ given a left- or right-tail violation, respectively:
\begin{equation}
\label{eqn:ES}
\rsES_{a_q, t}^{\rsTO} = \mathbb{E}{\left[X_t \middle| \rsVaRI_{a_q}^{\rsTO} = 1\right]} = \pm \frac{1}{a_q} \int_{\mp\infty}^{\rsVaR_{a_q, t}^{\rsTO}}{X' f_{X,t}{\left(X' \right)} d{X'}} ,
\end{equation}
where $\mathbb{E}{[.]}$ is the expectation operator and $f_{X,t} = d{F_{X,t}}/d{X}$ is the conditional probability density function (pdf) of $X_t$. Note that throughout this paper $F$ and $f$ are used to denote cumulative distribution functions and probability density functions, respectively. These are accompanied with subscripts to denote the variable or model associated with the distribution. Conditional distributions are also indexed in discrete time with the subscript $t$.

\subsection{\label{ssec:Hawkes}2T-POT Hawkes model}

\subsubsection{\label{sssec:HawkesTail}Hawkes exceedance model}

The two-tailed peaks-over-threshold (2T-POT) Hawkes model developed in \cite{Tomlinson2021} defines two sets of exceedance events in $X_{t}$. Left-tail exceedances events are defined as the values of $X_{t}$ that are less than left-tail exceedance threshold $u^{\rsTL}$; this series of events, which is indexed with $k$, is fully described by the series of left-tail excess magnitudes $\{M_{k}^{\rsTL}\} = \{ -(X_{t}-u^{\rsTL}) | -(X_{t}-u^{\rsTL})>0\}$ and arrival times $\{t_{k}^{\rsTL}\} = \{ t | -(X_{t}-u^{\rsTL})>0\}$. Similarly, right-tail exceedance events are defined with respect to the right-tail exceedance threshold $u^{\rsTR}$, and are fully specified by the series $\{M_{k}^{\rsTR}\}=\{+(X_{t}-u^{\rsTR}) | +(X_{t}-u^{\rsTR})>0\}$ and $\{t_{k}^{\rsTR}\}=\{t | +(X_{t}-u^{\rsTR})>0\}$. The thresholds $\vect{u}=\left[u^{\rsTL},u^{\rsTR}\right]^{\mathrm{T}}$ are defined symmetrically, so that number of exceedance events from each tail within the sample $X_{0:T}$ is asymptotically equal as the length of the sample $T$ tends to infinity. This is simply achieved by setting the thresholds equal to the value of mirrored in-sample quantiles, which is the typical approach to threshold selection in the financial POT Hawkes literature \citep{Chavez-Demoulin2005, Embrechts2011, Grothe2014, Ait-Sahalia2015, Gresnigt2015, Ait-Sahalia2015, BienBarkowska2020}. Here, these quantiles are specified by the threshold level $a_u \in [0,0.5]$, such that the left-tail threshold is equal to the in-sample $a_u$-quantile, $u^{\rsTL} = \hat{Q}_{a_u}(X_{0:T_{\mathrm{in}}})$, and the right-tail threshold is equal to its mirrored in-sample quantile, $u^{\rsTR} = \hat{Q}_{1-a_u}(X_{0:T_{\mathrm{in}}})$. 

In the most general description, the arrivals of left- and right-tail exceedance events are counted by two distinct point processes, which can be viewed as the components of the bivariate point process $\vect{N}{\left(t\right)} = [N^{\rsTL}{(t)}, N^{\rsTR}{(t)}]^{\mathrm{T}}$, such that
\begin{equation}
\label{eqn:dN}
d{N^{\rsTO}}{\left(t\right)} = \sum_{k}{\delta{\left(t-t_{k}^{\rsTO}\right)}} ,
\end{equation}
where $\delta{\left(t'\right)}$ is the Dirac delta function. The arrival rate at time $t$ of events within either point process is the conditional intensity for that process,
\begin{equation}
\label{eqn:lambda_defn}
\lambda^{\rsTO}{\left(t \middle| \mathcal{M}_t\right)} = \mathbb{E}{\left[ \frac{d{N^{\rsTO}}{\left(t\right)}}{dt} \middle| \mathcal{M}_t\right]} .
\end{equation}
The explicit time-dependence of $\lambda^{\rsTO}{\left(t \middle| \mathcal{M}_t\right)}$ specifies ${N^{\rsTO}}{\left(t\right)}$ to be \textit{inhomogeneous} point processes; Hawkes-type behavi\ENou{}r is specified by the conditional dependence on the event history up to the present time $t$, $\mathcal{M}_t = \mathcal{M}_{0:t} = \left\{\left(t_{k}^{\rsTO}, M_{k}^{\rsTO}\right): t_{k}^{\rsTO} < t\right\}$. If $N^{\rsTL}$ and $N^{\rsTR}$ are treated as distinct point processes, the conditional probability of a left-tail (right-tail) exceedance event occurring at time $t$ is
\inOneTwoColumn{
\begin{equation}
\label{eqn:p_bi}
p_{t}^{\rsTO} 
= \mathrm{Pr}{\left\{ \mathcal{I}\left[\mp \left(X_t - u^{\rsTO}\right)\right]=1 \middle| \mathcal{M}_t\right\}}
= \mathrm{Pr}{\left\{ d{N^{\rsTO}(t)} = 1 \middle| \mathcal{M}_t\right\}}
= 1 - \exp{\left[-\int_{t-1}^{t}{\lambda^{\rsTO}{\left(t' \middle| \mathcal{M}_t\right)} d{t'}}\right]} .
\end{equation}
}{
\begin{align}
\label{eqn:p_bi}
p_{t}^{\rsTO} 
&= \mathrm{Pr}{\left\{ \mathcal{I}\left[\mp \left(X_t - u^{\rsTO}\right)\right]=1 \middle| \mathcal{M}_t\right\}}
\nonumber\\
&= \mathrm{Pr}{\left\{ d{N^{\rsTO}(t)} = 1 \middle| \mathcal{M}_t\right\}}
\nonumber\\
&= 1 - \exp{\left[-\int_{t-1}^{t}{\lambda^{\rsTO}{\left(t' \middle| \mathcal{M}_t\right)} d{t'}}\right]} .
\end{align}
}
Note, however, that this is incompatible with the fact that the arrivals of these events within $X_t$ are mutually exclusive. Moreover, this treatment cannot natively forbid the case where the probability of an exceedance from either tail occurring at time $t$ is found to be greater than \ENone{}, i.e.\ENLA{} $p_{t}^{\rsTA} = p_{t}^{\rsTL} + p_{t}^{\rsTR} > 1$. It is established in \cite{Tomlinson2021} that, in the context of financial log-returns, these two requirements can be enforced at an insignificant cost to the goodness of fit by using the common intensity 2T-POT Hawkes model in which both types of extreme are counted within the same common point process $N^{\rsTA}{(t)}$, whose arrival rate is given by the one-dimensional common conditional intensity $\lambda^{\rsTA}$. Each exceedance event within $N^{\rsTA}{(t)}$ is then stochastically drawn from either tail upon arrival. Since the left- and right-tail thresholds are selected so that the average expected intensities of events from either tail, $a_{\lambda}^{\rsTO} \equiv \mathbb{E}{[\lambda^{\rsTO}]}$, are approximately equal ($a_{\lambda}^{\rsTL} = a_{\lambda}^{\rsTR} = a_{\lambda}^{\rsTA}/2$), the high correlation of left and right tail extremes \citep{Embrechts2011, Tomlinson2021} means that it can be assumed that each event in $N^{\rsTA}{(t)}$ is drawn from either tail with equal probability. Under this assumption, the conditional probability of a left-tail (right-tail) exceedance event occurring at time $t$ is
\inOneTwoColumn{
\begin{equation}
\label{eqn:p_ci}
p_{t}^{\rsTO}
= \frac{1}{2}\mathrm{Pr}{\left\{ d{N^{\rsTA}(t)} = 1 \middle| \mathcal{M}_t\right\}}
= \frac{1}{2}\left\{1 - \exp{\left[-\int_{t-1}^{t}{\lambda^{\rsTA}{\left(t' \middle| \mathcal{M}_t\right)} d{t'}}\right]} \right\} .
\end{equation}
}{
\begin{align}
\label{eqn:p_ci}
p_{t}^{\rsTO}
&= \frac{1}{2}\mathrm{Pr}{\left\{ d{N^{\rsTA}(t)} = 1 \middle| \mathcal{M}_t\right\}}
\nonumber\\
&= \frac{1}{2}\left\{1 - \exp{\left[-\int_{t-1}^{t}{\lambda^{\rsTA}{\left(t' \middle| \mathcal{M}_t\right)} d{t'}}\right]} \right\} .
\end{align}
}

\begin{figure*}
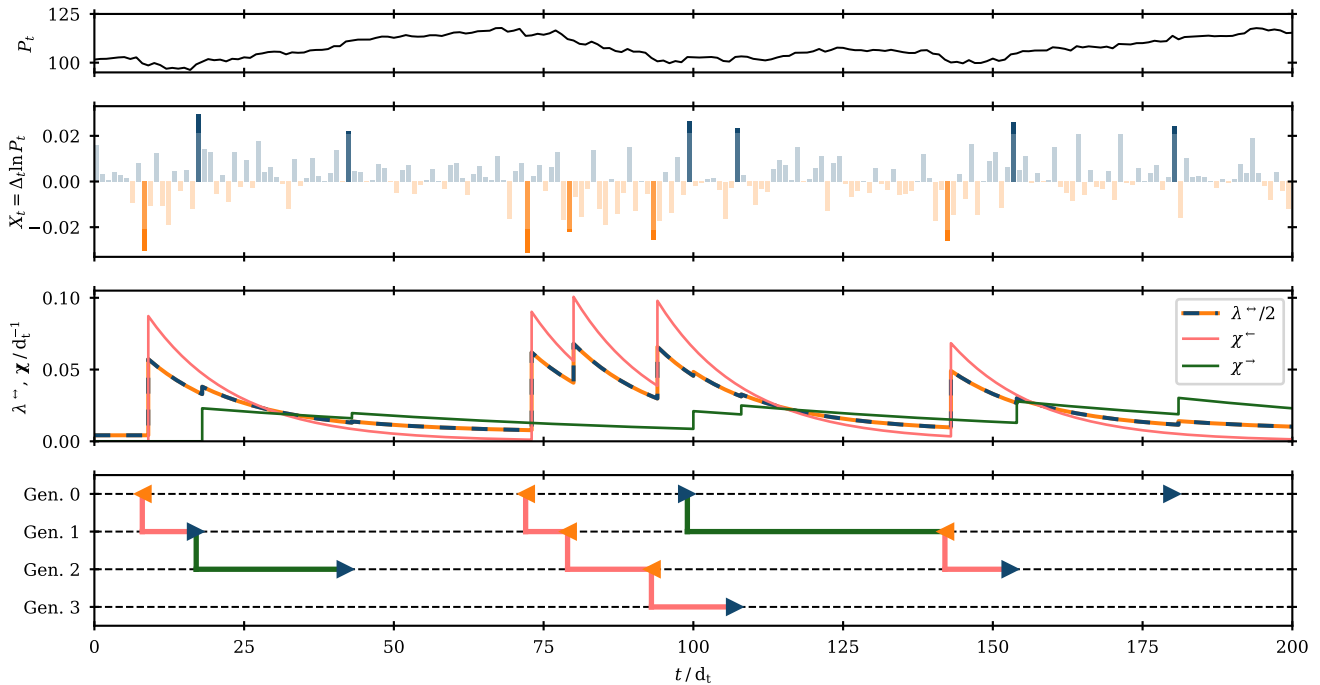

\includegraphics{\rsFolderSlash{\rsTickTrainfolder{SPX}}_ILL_branch_tt_\rsHwkMod-\rsHwkMode-mark-0025}
\caption{
\label{fig:ILL_branch_m1}
Illustration of the self-exciting 2T-POT Hawkes exceedance model as a branching process. The geometric random walk discrete time series $P_t$ (top panel) is transformed into the series of log-differences, $X_t = \Delta_{t}{\ln{P_{t}}} = \ln{P_{t}} - \ln{P_{t-1}} =\ln{P_{t}/P_{t-1}}$ (second panel from top), with threshold-exceeding ``extremes'' values marked in bolder shading. The 2T-POT Hawkes exceedance model (third panel) describes the common conditional intensity $\lambda^{\rsTA}$ (light orange and dark blue dashed) of left- and right-tail exceedance events as a linear sum of the endogenous excitements $\vect{\upchi}$ generated by the arrival of past left- (light red) and right-tail (dark green) exceedances. This may be understood as a branching process (fourth panel) in which the daughter events in generation $n+1$ are spawned from the endogenous intensity produced by mother events in generation $n$.
}
\end{figure*} 

The Hawkes-type arrival dynamics of the common intensity process are constructed through a constrained bivariate model\footnote{We note that when the thresholds are selected so that $a_{\lambda}^{\rsTL} = a_{\lambda}^{\rsTR}$, an unconstrained bivariate model of the form, $\vect{\uplambda} = \vect{\upmu} + \vect{\Gamma} \vect{\upchi}$, will tend to have equal rows in $\vect{\upmu}$ and the branching matrix $\vect{\Gamma}$, such that $\lambda^{\rsTL} \approx \lambda^{\rsTR}$. \cref{tab:p_LR_ci} shows that likelihood ratio test almost never finds a significant difference in goodness of fit between the full bivariate model [$H_{2,\mathrm{bi}}{\left(a_u\right)}$] and the common intensity model [$H_{2}{\left(a_u\right)}$], even in the out-of-sample period where the $a_{\lambda}^{\rsTL} = a_{\lambda}^{\rsTR}$ condition is not strictly guaranteed by the thresholds (which are set to mirrored in-sample quantiles).} that still allows for asymmetric self- and cross-excitement between asymmetric tails. This takes the form
\begin{equation}
\label{eqn:common_intensity_model}
{\lambda^{\rsTA}}{\left(t \middle| \vect{\uptheta}_{u}; \mathcal{M}_t\right)} = \mu^{\rsTA} + {\vect{\upgamma}^{\rsTA}}^{\mathrm{T}} \vect{\upchi}{\left(t \middle| \vect{\uptheta}_{u}; \mathcal{M}_t\right)} ,
\end{equation}
where $\vect{\uptheta}_{u}$ is the parameter vector for the Hawkes exceedance model, $\mu^{\rsTA}$ is the constant exogenous background intensity for the common arrival process, $\vect{\upgamma}^{\rsTA} \equiv [\gamma^{\rsTA\rsTL}, \gamma^{\rsTA\rsTR}]^{\mathrm{T}}$ is the branching vector, and $\vect{\upchi}{\left(t \middle| \vect{\uptheta}_{u}; \mathcal{M}_t\right)} \equiv [\chi^{\rsTL}, \chi^{\rsTR}]^{\mathrm{T}}$ is the vector of endogenous excitements generated by the arrivals of left- and right-tail exceedance events. 

As is illustrated in \cref{fig:ILL_branch_m1}, the self-exciting dynamics of the Hawkes process can be understood as a branching process, in which \textit{daughter} events are triggered by the additional endogenous intensity produced by the arrival of prior \textit{mother} events. $\vect{\upgamma}^{\rsTA}$ is called the branching vector, because $\gamma^{\rsTA\rsTO}$ is the mean number of daughter events in $N^{\rsTA}$ that are triggered by a mother event in $N^{\rsTO}$. This is so, because the endogenous excitement $\chi^{\rsTO}$ is normali\ENz{}ed, such that the expected lifetime contribution of each event in $N^{\rsTO}$ to $\chi^{\rsTO}$ is \ENone{}. This normali\ENz{}ation also guarantees that the model is uniquely fitted. The process is sub\ENhyph{}critical (i.e.\ENLA{} non\ENhyphNon{}explosive) provided the aggregate branching ratio $(\gamma^{\rsTA\rsTL} + \gamma^{\rsTA\rsTR})/2$ is less than \ENone{} \citep{Wheatley2019}.

The components $\chi^{\rsTO}$ of the endogenous excitement vector $\vect{\upchi}$ are the sums of contributions from all past events in each tail,
\begin{equation} 
\label{eqn:1DHawkesEndo}
\chi^{\rsTO}{\left(t \middle| \vect{\uptheta}_{u}; \mathcal{M}_t\right)} = \sum_{k:t_{k}^{\rsTO}<t}{\phi^{\rsTO}{\left(t-t_{k}^{\rsTO}\right)\kappa^{\rsTO}{\left(M_{k}^{\rsTO} \middle| t_{k}^{\rsTO}\right)}}} ,
\end{equation}
where, for each component, the decay kernel $\phi^{\rsTO}$ is a monotonically decreasing function of the time between the arrival of the past event, $t_{k}^{\rsTO}$, and the present, $t$. The 2T-POT Hawkes model was conceived as a parametric model: for the sake of parsimony when describing multiple tails independently, the decay kernels -- along with the other functions on which the endogenous excitement depends -- are described with parametric functions adapted from previous financial POT literature \citep{Grothe2014, Gresnigt2015}. The decay kernel has previously been taken as either an exponential or power law decay \citep{Gresnigt2015}. \cite{Tomlinson2021} used an exponential decay, which is scaled differently for the left and right tails according to the decay vector $\vect{\upbeta} = \left[\beta^{\rsTL}, \beta^{\rsTR}\right]^{\mathrm{T}}$, such that
\begin{equation} 
\label{eqn:phi_exp}
\phi^{\rsTO}{\left(t'\right)} = \beta^{\rsTO}e^{-\beta^{\rsTO} t'} .
\end{equation}
The advantages of this choice are that \cref{eqn:1DHawkesEndo} can be recast in Markov form,
\begin{equation} 
\label{eqn:1DHawkesEndo_Markov}
d{\chi^{\rsTO}} = \beta^{\rsTO} \left[- \chi^{\rsTO} d{t} + \kappa^{\rsTO}{\left(M_{k}^{\rsTO} \middle| t_{k}^{\rsTO}\right)} d{N^{\rsTO}} \right] ,
\end{equation}
and the decay vector $\vect{\beta}$ can be used to infer the characteristic timescales over which daughter events are triggered by mother events from either tail.

The conditional impact function $\kappa^{\rsTO}{(M|t)}$ is a monotonically increasing function of the excess magnitude $M$. Following the approach of \cite{Grothe2014, Gresnigt2015}, this is defined so that the intensity jump from the exceedance event arriving at time $t_{k}^{\rsTO}$ is determined by the contemporaneous value of the conditional cumulative distribution function of excess magnitudes for that tail
\begin{equation} 
\label{eqn:kappa}
\kappa^{\rsTO}{\left(M \middle| t\right)} = \frac{1 - \alpha^{\rsTO} \ln{\left[1 - F_{M,t}^{\rsDistGP,\rsTO}{\left(M\right)} \right]}}{1 + \alpha^{\rsTO}},
\end{equation}
where the mark parameter $\alpha^{\rsTO} \geq 0$. When $\alpha^{\rsTO} \neq 0$, larger magnitude events produce greater jumps in the excitement. This reduces sensitivity on the choice of threshold value $u^{\rsTO}$, since $\kappa^{\rsTO}{(M|t)} \rightarrow (1+\alpha^{\rsTO})^{-1}$ as $M \rightarrow 0$. Conversely, when $\alpha^{\rsTO}=0$, $\kappa_{t}^{\rsTO}$ becomes unity and an \textit{unmarked} Hawkes process -- in which $\chi^{\rsTO}$ is independent of the magnitudes of past events -- is recovered. \cref{eqn:kappa} is specified so that $\mathbb{E}{\left[\kappa^{\rsTO}{\left(M|t\right)}\right]} \equiv 1$ for all values of $\alpha^{\rsTO}$ and $t$. Also, note that throughout this paper superscript calligraphic letters symboli\ENz{}e parametric probability distributions: $\rsDistNorm$ for the normal distribution, $\rsDistT$ for the Student-$t$ distribution, and $\rsDistGP$ for the generali\ENz{}ed Pareto distribution. Hence, $F_{M,t}^{\rsDistGP,\rsTO}$ in \cref{eqn:kappa} denotes that the conditional cdf of excess magnitudes from either tail is described by the GP distribution. 

In EVT methods, the tail is almost always described by the GP distribution. This follows from the Gnedenko-Pickands-Balkema-de Haan (GPBH) theorem, which states that the GP distribution is the limiting distribution for threshold excesses \citep{Pickands1975, Balkema1974, Sornette2006}. Accordingly, we assume that the excess magnitudes are distributed according to a conditional GP distribution, as specified by the cdf
\begin{equation} 
\label{eqn:F_GPD}
F_{M,t}^{\rsDistGP,\rsTO}{\left(M\right)} = 
\begin{cases}
1 - {\left[1 + \xi^{\rsTO}M/\sigma_{t}^{\rsTO}\right]}^{-1/\xi^{\rsTO}} , & \xi_{\rsTO} \neq 0 , \\
1 - \exp{\left[- M/\sigma_{t}^{\rsTO}\right]} , & \xi^{\rsTO} = 0 ,
\end{cases}
\end{equation}
where the shape parameter $\xi^{\rsTO}$ can specify a range of tail heaviness over three distinct phases: from the finite decay of the Weibull distribution ($\xi^{\rsTO} < 0$), through the exponential decay of the Gumbel distribution ($\xi^{\rsTO} = 0$), to the increasingly leptokurtic power-law decay of the Fr\'{e}chet distribution ($\xi^{\rsTO} > 0$) \citep{Grothe2014, Sornette2006}. Conditional dependence on the endogenous (i.e.\ENLA{} non-background) intensity of the Hawkes process is introduced via the conditional scale parameter
\begin{equation} 
\label{eqn:sigma}
\sigma_{t}^{\rsTO} = \varsigma^{\rsTO} + \eta^{\rsTO} \left[\lambda^{\rsTA}{\left(t\right)} - \mu^{\rsTA} \right]/2 .
\end{equation}
Thus, when $\eta^{\rsTO} > 0$, larger magnitude events become more likely when the conditional intensity of exceedances is high, as is generally observed in financial returns \citep{Cont2001}. Conversely, when $\eta^{\rsTO} = 0$ the excess magnitudes are drawn from an unconditional GP distribution with a fixed scale parameter $\sigma_{t}^{\rsTO} = \varsigma^{\rsTO}$.

In this paper, we greatly expand the application of the 2T-POT Hawkes model to historic market data: in \cite{Tomlinson2021}, the model was fitted to a single data sample (SPX) at a single threshold level, $a_u=0.025$; in \cref{sec:Data} of this paper, we not only expand this to 6 data samples, but, uniquely within the financial POT Hawkes literature, we also fit the model to each sample at 20 different values of the threshold level, $a_u = 0.0125 k_u$, where $k_u \in \mathbb{Z} \cup [1,20]$. Given that the computational cost of calibrating the exceedance model scales with $a_{u}$, the 120 independent calibrations in this paper are approximately 630 times more expensive in total than the single calibration performed in \cite{Tomlinson2021}. This greatly expanded application of the 2T-POT Hawkes model required improvements to the speed and reliability of the optimi\ENz{}ation procedure. We achieved this achieved by reparametri\ENz{}ing the exceedance model so that expected average intensity $a_{\lambda}^{\rsTA} = \mathbb{E}{[\lambda^{\rsTA}]}$ replaced the background intensity $\mu^{\rsTA}$ as a fitting parameter, with the latter then calculated as
\begin{equation}
\label{eqn:Hawkes_mu_constraint}
\mu^{\rsTA} = \left[2 - \left(\gamma^{\rsTA\rsTL} + \gamma^{\rsTA\rsTR}\right)\right] a_{\lambda}^{\rsTA} ,
\end{equation}
as is derived in \cref{sec:reparam}. It is more efficient to use $a_{\lambda}^{\rsTA}$ as a fitting parameter instead of $\mu^{\rsTA}$, since the former is orthogonal to $\vect{\upgamma}^{\rsTA}$. When this updated model was used to reproduce the single calibration from \cite{Tomlinson2021}, we found that the maximum likelihood (ML) optimi\ENz{}ation time under the SLSQP method in SciPy \citep{Nocedal2006c18} was reduced by 53\%. More significantly, this modification also greatly reduced the number of failed optimi\ENz{}ations in the expanded set of applications performed in this paper. Indeed, we emphasise that the large-scale analysis in \cref{sec:Data,sec:CT_QE} was only made feasible because of this reparametri\ENz{}ation. Even greater efficiency was achieved by noting the relationship between the expected average intensity $a_{\lambda}^{\rsTA}$ and the threshold level $a_u$. Because the thresholds $\vect{u}$ are set equal to the $(a_u, 1-a_u)$ in-sample quantiles, it follows that the average intensity $a_{\lambda}^{\rsTA}$ in the in-sample period is asymptotically equal to $2 a_{u} \rsUdt$, where $\rsUdt$ is the unit measuring one step in discrete time (in the case of daily log-returns, $\rsUdt$ denotes trading days), as the size of the in-sample period $T_{\mathrm{in}}$ tends to infinity. Thus, $a_{\lambda}^{\rsTA} = 2 a_u \rsUdt$ can be used either as an initial value in estimation or used as a fixed constraint. It is shown by likelihood ratio tests in \cref{sec:model_select} that the constraint achieves a parameter reduction of \ENone{} at negligible cost to the goodness of fit; accordingly, this constraint is enforced for all calibrations in this paper. We also note that the constraint reduced the total optimi\ENz{}ation time for this paper by 12\%. Having demonstrated these significant gains in optimi\ENz{}ation speed and reliability, we strongly recommend that our reparameteri\ENz{}ation is applied to all other Hawkes models that use the background intensity as a fitting parameter.

The common intensity 2T-POT Hawkes exceedance model is fully specified by the set of parameters, $\vect{\uptheta}_{u} =\{
	\vect{\upgamma}^{\rsTA},
	\vect{\upbeta},
	\vect{\upxi},
	\vect{\varsigma},
	\vect{\upeta},
	\vect{\alpha};
	a_u
\}$, where vector quantities are of the form, $\vect{\upbeta} \equiv [\beta^{\rsTL}, \beta^{\rsTR}]^{\mathrm{T}}$. This model is hereafter denoted as $H_2{(a_u)}$. Note that, if all parameters in $\vect{\uptheta}_{u}$ are constrained to be symmetric (so that the left- and right-tail components of all vector parameters are equal), then the common intensity 2T-POT model is equivalent to the classical single-tailed peaks-over-threshold Hawkes model applied to the absolute values of a copy of the original time series that is cent\ENere{}d on the mid-point between the thresholds. That is, the set of absolute exceedances $\{M_{k}^{\rsTA}\}=\{|X_{t} - (u_{\rsTR} + u_{\rsTL})/2| - u^{\rsTA} > 0\}$, where $u^{\rsTA} = (u^{\rsTR} - u^{\rsTL})/2$, is a union of $\{M_{k}^{\rsTL}\}$ and $\{M_{k}^{\rsTR}\}$, and a univariate Hawkes model applied to this exceedance series describes equal self- and cross-excitations between left and right tails that are symmetric in all properties. Hereafter, this is referred to as the symmetric 2T-POT Hawkes exceedance model and is denoted as $H_1{(a_u)}$.

\subsubsection{\label{sssec:HawkesBulk}Subordinate bulk distribution}

It is simple to calculate left-tail (right-tail) conditional quantile at the coverage level $a_q$ over the holding period $t-1$ to $t$ (i.e.\ENLA{} to calculate $\rsVaR_{a_q, t}^{\rsTO}$) using the 2T-POT Hawkes exceedance model provided that the conditional probability of a left-tail (right-tail) exceedance estimated by the model is greater than or equal to the coverage level, i.e.\ENLA{} $p_{t}^{\rsTO} \geq a_q$. In this case, $\rsVaR_{a_q, t}^{\rsTO}$ will lie within the GP tail distribution, $F_{M,t}^{\rsDistGP,\rsTO}$, specified by the exceedance model. Thus,
\inOneTwoColumn{
\begin{align}
a_q 
&= \mathrm{Pr}{\left\{ \mathcal{I}\left[\mp \left(X_t - u^{\rsTO}\right)\right]=1 \middle| \mathcal{M}_t\right\}}
\nonumber\\
&\times \mathrm{Pr}{\left\{ \mathcal{I}\left[\mp \left(X_t - \rsVaR_{a_q}^{\rsTO}\right)\right]=1 \middle|\mathcal{I}\left[\mp \left(X_t - u^{\rsTO}\right)\right]=1; \mathcal{M}_t\right\}}
\nonumber\\
&= p_{t}^{\rsTO} \left[1 \mp \xi^{\rsTO} \frac{\rsVaR_{a_q, t}^{\rsTO} - u^{\rsTO}}{\sigma_{t}^{\rsTO}}\right]^{-1/\xi^{\rsTO}} .
\label{eqn:2T-POT_Q_a_q}
\end{align}
}{
\begin{align}
a_q 
&= \mathrm{Pr}{\left\{ \mathcal{I}\left[\mp \left(X_t - u^{\rsTO}\right)\right]=1 \middle| \mathcal{M}_t\right\}}
\nonumber\\
&\times \mathrm{Pr}{\left\{ \mathcal{I}\left[\mp \left(X_t - \rsVaR_{a_q}^{\rsTO}\right)\right]=1 \middle|\mathcal{I}\left[\mp \left(X_t - u^{\rsTO}\right)\right]=1; \mathcal{M}_t\right\}}
\nonumber\\
&= p_{t}^{\rsTO} \left[1 \mp \xi^{\rsTO} \frac{\rsVaR_{a_q, t}^{\rsTO} - u^{\rsTO}}{\sigma_{t}^{\rsTO}}\right]^{-1/\xi^{\rsTO}} .
\label{eqn:2T-POT_Q_a_q}
\end{align}
}
Hence,
\begin{equation}
\label{eqn:2T-POT_Q}
\left( \rsVaR_{a_q, t}^{\rsTO} | p_{t}^{\rsTO} \geq a_q \right) = u^{\rsTO} \mp \left[\left(\frac{a_q}{p_{t}^{\rsTO}}\right)^{-\xi^{\rsTO}} - 1\right]\frac{\sigma_{t}^{\rsTO}}{\xi^{\rsTO}} .
\end{equation}
The conditional violation expectation over the same holding period can then be calculated as
\begin{equation}
\label{eqn:2T-POT_E}
\left( \rsES_{a_q, t}^{\rsTO} | p_{t}^{\rsTO} \geq a_q \right) = \rsVaR_{a_q, t}^{\rsTO} \pm \frac{\xi^{\rsTO}(\rsVaR_{a_q, t}^{\rsTO} - u^{\rsTO}) - \sigma_{t}^{\rsTO}}{1 - \xi^{\rsTO}} .
\end{equation}

If, however, the left-tail (right-tail) exceedance probability is less than the coverage level (i.e.\ENLA{} $p^{\rsTO, t} < a_q$), then the hypothetical conditional quantile would lie 
between the two thresholds -- within the bulk of log-returns not supported by either the left- or right-tail exceedance distributions of the 2T-POT Hawkes model. It is therefore desirable to extend the support of the 2T-POT Hawkes model to the full distribution of $X_t$. We do this by incorporating it as the jump process within a jump-diffusion type model where the intra\ENhyph{}threshold bulk is supported by a supplementary conditional distribution that describes diffusion. Even when solely concerned with the forecasting of extremes, the distribution of the intra\ENhyph{}threshold bulk becomes increasingly pertinent as the coverage level $a_q$ approaches the threshold level $a_u$ from below, and even more so if forecasts are to be made over more than a single time step. This extension was essential for this paper because the backtesting methods used in \cref{sec:CT_QE} require that $\rsVaR_{a_q, t}^{\rsTO}$ and $\rsES_{a_q, t}^{\rsTO}$ are defined at all values of $t$, and this can only be guaranteed when the full distribution of $X_t$ is supported. 

Because we are interested in the Hawkes process as a pure conditional EVT model and how this compares to the conditional volatility mechanism described by the GARCH-EVT model, we take the unusual step of subordinating the diffusion model (responsible for the bulk) to the jump model (generating the extreme events). Specifically, we introduce an intra\ENhyph{}threshold bulk distribution that is transformed by parameters that are conditional upon the intensity of the Hawkes exceedance process. This subordinate bulk distribution is described by the conditional cdf $F_{B,t}^{\rsDistD}$, where the superscript $\rsDistD$ is replaced by the symbol of the specified parametric distribution ($\rsDistNorm$ for the normal distribution and $\rsDistT$ for the Student-$t$ distribution). $F_{B,t}^{\rsDistD}$ is transformed by the conditional location and scale parameters, $m_t$ and $s_t$, so that its value at the thresholds $\vect{u}$ matches the probability of an exceedance event at time $t$ as determined by the Hawkes arrival process. These two constrains are expressed by the equations,
\begin{subequations}
\begin{align}
\label{eqn:F_norm_L}
F_{B,t}^{\rsDistD}{\left(u^{\rsTL} \middle| m_t, s_t^2; \vect{\uptheta}_{B}^{D} \right)} &= p_{t}^{\rsTL} , \\
\label{eqn:F_norm_R}
1 - F_{B,t}^{\rsDistD}{\left(u^{\rsTR} \middle| m_t, s_t^2; \vect{\uptheta}_{B}^{D}\right)} &= p_{t}^{\rsTR} .
\end{align}
\end{subequations}
where $\vect{\uptheta}_{B}^{D}$ is a vector containing the additional unconstrained parameters of the chosen parametric distribution $\rsDistD$. Here, it becomes essential that the condition, $p_{t}^{\rsTL} + p_{t}^{\rsTR} \leq 1$ is strictly enforced by the use of a common conditional intensity as described in \cref{eqn:p_ci}. In contrast, the bivariate model described in \cref{eqn:p_bi} does not enforce this condition, and so it will attribute negative probability mass to the intra-threshold bulk when the components of the bivariate conditional intensity $\lambda^{\rsTO}$ are sufficiently high. 

The conditional pdf for $X_{t}$ at time $t$ is then a weighted piecewise union of the bulk and tail distributions: 
\begin{equation}
\label{eqn:f_norm_Hawkes}
f_{H,t}^{\rsDistD}{\left(X\right)} = 
\begin{cases}
p_{t}^{\rsTL} f_{M,t}^{\rsDistGP,\rsTL}{\left[-\left(X - u^{\rsTL}\right)\right]}, & X < u^{\rsTL}, \\
f_{B,t}^{\rsDistD}{\left(X \middle| m_t, s_t\right)}, & u^{\rsTL} \leq X \leq u^{\rsTR}, \\
p_{t}^{\rsTR} f_{M,t}^{\rsDistGP,\rsTR}{\left[+\left(X - u^{\rsTR} \right)\right]}, & X > u^{\rsTR}.\\
\end{cases}
\end{equation}
The fully supported 2T-POT Hawkes model at threshold level $a_u$ is denoted as $H_2^{\rsDistD}{(a_u)}$. The equivalent fully supported symmetric 2T-POT Hawkes model is denoted as $H_{1}^{\rsDistD}{(a_u)}$.

\subsection{\label{ssec:GARCH}GARCH-EVT model}

\subsubsection{\label{sssec:GARCH}GARCH}

In this paper, the 2T-POT Hawkes model is compared with the family of generali\ENz{}ed autoregressive conditional heteroscedasticity (GARCH) models. These are the standard reduced form models for log-returns in industry and they are a ubiquitous baseline in the financial risk literature \citep{Ruppert2015}. GARCH models have earned this status due to their parsimonious description of two key styli\ENz{}ed facts of financial returns: volatility clustering, which states that the standard deviation of log-returns (known as the volatility) is non\ENhyphNon{}constant and exhibits significant positive autocorrelation, and the leverage effect, which states that volatility increases when returns become more negative \citep{Cont2001, Ruppert2015}. GARCH models describe a conditional volatility process in which a real univariate discrete time series is generated as
\begin{equation}
\label{eqn:GARCH_epsilon}
X_t = \mu + \sigma_t \epsilon_{t}^{\rsDistD} ,
\end{equation}
where $\mu$ is the unconditional mean, $\sigma_t$ is the conditional volatility, and $\epsilon_{t}^{\rsDistD}$ are i.i.d.\@ random innovations drawn from a parametric distribution $\rsDistD$, typically with zero mean and unit variance. The $\mathrm{GARCH}{(p,r,q)}$ model describes the conditional variance $\sigma_t^2$ with an autoregressive moving average (ARMA) process
\inOneTwoColumn{
\begin{equation}
\label{eqn:GARCH}
\sigma_t^2 = \omega 
+ \sum_{i=1}^{p}{\alpha_{i}{\left(\sigma_{t-i}\epsilon_{t-i}\right)}^2} + \sum_{j=1}^{q}{\beta_{j}\sigma_{t-j}^2}
+ \sum_{k=1}^{r}{\gamma_{k}{\left(\sigma_{t-k}\epsilon_{t-k}\right)}^2 \mathcal{I}{\left[-\epsilon_{t-k}\right]}}
,
\end{equation}
}{
\begin{align}
\sigma_t^2 = \omega 
&+ \sum_{i=1}^{p}{\alpha_{i}{\left(\sigma_{t-i}\epsilon_{t-i}\right)}^2} + \sum_{j=1}^{q}{\beta_{j}\sigma_{t-j}^2}
\nonumber\\
&+ \sum_{k=1}^{r}{\gamma_{k}{\left(\sigma_{t-k}\epsilon_{t-k}\right)}^2 \mathcal{I}{\left[-\epsilon_{t-k}\right]}}
,
\label{eqn:GARCH}
\end{align}
}
where $\omega$ is the minimum conditional variance and $\{\alpha_i, \beta_j, \gamma_k\}$ are the ARCH coefficients. When $r > 0$, \cref{eqn:GARCH} is a GJR-GARCH model \citep{Glosten1993} that accounts for the leverage effect -- interpreted as an asymmetric impact of negative log-returns on volatility -- through the Heaviside step function. 

In this paper, we consider the $\mathrm{GARCH}{(p=1,r,q=1)}$ model with ($r=1$) and without ($r=0$) the leverage effect. For the innovation distribution, we consider the unit normal distribution ($\rsDistD = \rsDistNorm_{0,1}$) and the unit Student-$t$ distribution with $\nu$ degrees of freedom ($\rsDistD = \rsDistT_{0,1, \nu}$). Hereafter, we label these specifications of the GARCH model as ${\mathrm{G}_{r}^{\rsDistD}{(0)}}$.

\subsubsection{\label{sssec:GARCHEVT}GARCH-EVT}

\begin{table*}
\caption{\label{tab:data_stat}
Summary statistics for the series of daily log-returns of the six large cap international equity indices. For each sample of log-returns $X_t$, the number of observations $T$, mean $\bar{X}$, standard deviation $\sigma_X$, median $\hat{Q}_{0.5}{(X)}$, and median absolute deviation $\mathrm{MAD}_X$ are given.}
\FTABdatastat
\end{table*}

While the standard GARCH models specified in \cref{sssec:GARCH} provide the typical benchmark for forecasting log-returns, we are primarily interested in comparing the 2T-POT Hawkes model with the GARCH-EVT model \citep{McNeil2000, Echaust2020}. This model is constructed by appending GP tails to the distribution of innovations, so that the (unconditional) pdf for $\epsilon_t$ now takes the form
\begin{equation}
\label{eqn:f_norm_GARCH}
f_{\epsilon}^{\rsDistD}{\left(\epsilon\right)} = 
\begin{cases}
a_u f_{M_{\epsilon}}^{\rsDistGP,\rsTL}{\left(-\left\{\epsilon - u_{\epsilon}^{\rsTL} \right\}\right)} , & \epsilon_t < u_{\epsilon}^{\rsTL} ,\\
f_{B_{\epsilon}}^{\rsDistD}{\left(\epsilon\right)} , & u_{\epsilon}^{\rsTL} \leq \epsilon_t \leq u_{\epsilon}^{\rsTR} , \\
a_u f_{M_{\epsilon}}^{\rsDistGP,\rsTR}{\left(+\left\{\epsilon - u_{\epsilon}^{\rsTR} \right\}\right)} , & \epsilon_t > u_{\epsilon}^{\rsTR} ,\\
\end{cases} 
\end{equation}
where $f_{B_{\epsilon}}^{D}$ is the pdf of a continuous parametric distribution of zero mean and unit variance\footnote{Note that $f_{\epsilon}^{\rsDistD}$ will have a different variance compared with $f_{B_{\epsilon}}^{\rsDistD}$.} and the GP tail distributions $f_{M_{\epsilon}}^{\rsDistGP,\rsTO}$ are parameteri\ENz{}ed equivalently to \cref{eqn:F_GPD} except that the scale parameter remains constant and is therefore denoted as $\varsigma_{\epsilon}^{\rsTO}$. Thus,
\begin{equation} 
\label{eqn:F_GPD_G}
f_{M_{\epsilon}}^{\rsDistGP,\rsTO}{\left(M_{\epsilon}\right)} = 
\begin{cases}
1 - {\left[1 + \xi_{\epsilon}^{\rsTO}M_{\epsilon}/\varsigma_{\epsilon}^{\rsTO}\right]}^{-1/\xi_{\epsilon}^{\rsTO}} , & \xi_{\epsilon}^{\rsTO} \neq 0 , \\
1 - \exp{\left[- M_{\epsilon}/\varsigma_{\epsilon}^{\rsTO}\right]} , & \xi_{\epsilon}^{\rsTO} = 0 .
\end{cases}
\end{equation}
The innovation thresholds are set according to the threshold level $a_u$, such that $u_{\epsilon}^{\rsTL} \equiv {F_{B_{\epsilon}}^{D}}^{-1}{\left(a_u\right)}$ and $u_{\epsilon}^{\rsTR} \equiv {F_{B_{\epsilon}}^{D}}^{-1}{\left(a_u\right)}$. Thus, when $a_u=0$, the standard (i.e.\ENLA{} non-EVT) GARCH models are recovered. Accordingly, the GARCH-EVT model is labe\ENll{}ed as ${\mathrm{G}_{r}^{\rsDistD}{(a_u)}}$. Estimates for the innovation tail parameters, $\hat{\vect{\upxi}}_{\epsilon}$ and $\hat{\vect{\varsigma}}_{\epsilon}$, are obtained by ML estimation over the estimated innovations $\hat{\epsilon_t} = {(X_t - \hat{\mu})/\hat{\sigma}_t}$ \citep{McNeil2000}. Note that we use hat accents to denote values that are estimated or are derived from estimates.

As well as being one of the best performing univariate risk models within the financial risk literature, the GARCH-EVT model represents an alternative conditional EVT approach compared with the 2T-POT Hawkes model. In the latter, the dynamics are solely influenced by the threshold exceeding events. Conversely, the dynamic properties of the former are functions of the conditional volatility $\sigma_{t}$, which, as per \cref{eqn:GARCH}, is influenced by each innovation in $\epsilon_t$ in proportion to their magnitude. Thus, while the 2T-POT Hawkes model faithfully extends the foundational principle of EVT -- that `extreme events should speak for themselves' -- to the conditional case, the GARCH-EVT model represents a compromise. The comparison of the forecasting accuracy of these two models in \cref{sec:CT_QE} therefore acts as comparison between these two distinct views of how extreme events are generated.

\section{\label{sec:Data}Data and empirical study}

\subsection{\label{ssec:Data_desc}Data description}

The models specified in \cref{sec:Model} are applied to the daily log-returns of six international large cap equity indices: the S\&P 500 (SPX) and Dow Jones Industrial Average (DJI) from the U.S.A., the DAX 30 (DAX) of Germany, the CAC 40 (CAC) of France, the Nikkei 225 (NKX) of Japan, and the Hang Seng index (HSI) of Hong Kong. These series are widely investigated financial benchmarks that are often perceived as proxies for the broader equity market in the world's three major financial cent\ENre{}s: North America, Western Europe, and East Asia. We specifically use the daily log-returns over the period beginning \rsDateTrainStart{}\footnote{The official base date for both the DAX 30 and CAC 40 indices is 1987-12-31; however, both of these series can be extended backwards using their direct predecessors: the former DAX index and the Insee de la Bourse de Paris, respectively.} and ending \rsDateAllEnd{}. This is divided into an in-sample training period (\rsDateTrainStart{} to \rsDateTrainEnd{}) used to calibrate the models in this section and an out-of-sample period (\rsDateTrainEnd{} to \rsDateAllEnd{}) used to backtest the next step ahead forecasts of $\hat{\rsVaR}_{a_q,t}^{\rsTO}$ and $\hat{\rsES}_{a_q,t}^{\rsTO}$ in \cref{sec:CT_QE}. The in-sample period is a forty-year span that includes among other notable episodes the 1987 Black Monday crash, the 1997-8 Asian Financial Crisis, the 2000-2002 Tech Bubble crash, and the 2007-8 Global Financial Crisis. The out-of-sample period encompasses more than 92 months of financial data and includes the severe fluctuations caused by the outbreak of the Covid-19 pandemic in March 2020 and by the Russian invasion of Ukraine in February 2022. The data were sourced from the stooq.pl online database \citep{Stooq2021}; summary statistics are provided in \cref{tab:data_stat}. We note that, for both the in- and out-of-sample periods, the six series are not of the exact same length $T$ due to idiosyncratic holidays and suspensions.

\subsection{\label{ssec:Estimation}Threshold selection and in-sample calibration}

\begin{figure*}
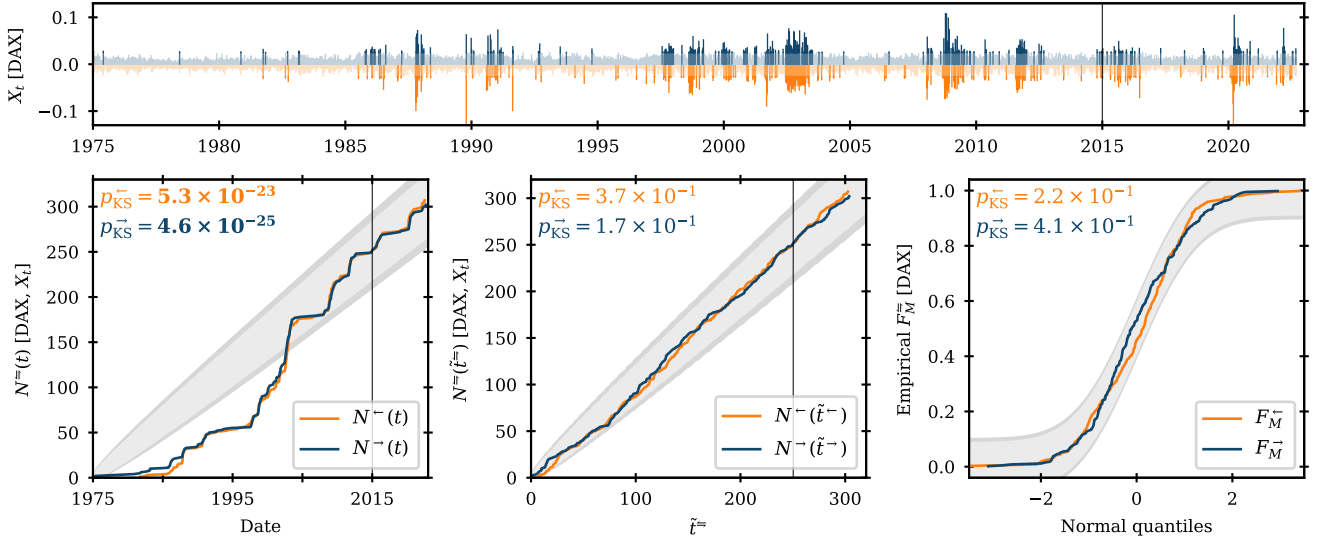

\includegraphics{\rsFolderSlash{\rsTickAllfolder{DAX}}_tick_tt_\rsHwkMod-\rsHwkMode-mark-0025}
\caption{
\label{fig:DAX_tick_m1}
$H_{2}{(a_u=0.025)}$ Hawkes exceedance model fitted to the DAX daily log-returns (top panel). The bottom panels show the KS test performed on the in-sample arrival process in time (bottom-left panel, $\mathcal{H}_0$: exceedance arrivals are Poisson in time $t$), the in-sample arrivals process in residual time (bottom-middle panel, $\mathcal{H}_0$: exceedance arrivals are unit Poisson in residual time $\tilde{t}^{\rsTO}$), and the residual excess magnitudes (bottom-right panel, $\mathcal{H}_0$: residual excess magnitudes $\tilde{m}_{k}^{\rsTO}$ are unit exponential random variables): the gr\ENe{}y shaded areas show the 95\% (lighter) and 99\% (darker) KS confidence intervals; the KS $p$-values for the left- (light orange) and right-tail (dark blue) processes are also shown, with rejections at the 95\% confidence level highlighted in bold. The vertical black lines mark the end of the in-sample period. Equivalents of this figure for the other indices are included in the supplementary material.
}
\end{figure*}

As is discussed in \cref{sssec:HawkesTail}, the exceedance thresholds $\vect{u}$ of the POT Hawkes models are determined by the choice of threshold level $a_u$, which specifies the quantiles at which the thresholds are set and is regarded as a tuneable parameter. In unconditional EVT applications, where an unconditional GP distribution is fitted to a stationary tail, the only concern with the threshold is that it determines the sample of excess magnitudes $M_{k}^{\rsTO}$. Threshold selection is therefore treated as a classic bias-variance trade\ENhyph{}off: a less extreme threshold (i.e.\ENLA{} higher threshold level $a_u$) ensures a greater number of observations in the tail; reducing noise and improving the stability of parameter estimates. However, the asymptotic nature of the GPDH theorem means that the exceedance distribution is better approximated by the GP distribution as the threshold becomes more extreme (i.e.\ENLA{} as the threshold level $a_u$ is lowered to \ENzero{}). Diagnostic tools exist for the latter issue \citep{Scarrott2012}, and these are often used to inform threshold selection in this time homogeneous case: for instance, in the construction of the GARCH-EVT model when fitting GP tails to the estimated innovations $\hat{\epsilon}_{t}$ \citep{Echaust2020}. However, the introduction of Hawkes-type arrival dynamics adds a second layer of importance to threshold selection, because this also determines the sample of arrival times $t_{k}^{\rsTO}$. This not only effects the expected distribution of the conditional intensity $\lambda^{\rsTA}$, but also distribution in time of the threshold exceeding events that provide information to the Hawkes process. While these additional effects invalidate the threshold selection procedures used in the stationary case, they may at the same time introduce distinct phases along $a_u$ in which the 2T-POT Hawkes model identifies signals from the data generating mechanisms of the underlying system. This could be evidenced through a phase transition along $a_u$ in the fitted parameters or in the forecasting accuracy, either of which could provide a physically meaningful definition of an extreme event in the combined context of arrivals and magnitudes. To explore this, we calibrate the threshold-based models across a wide range of threshold levels: $a_u = 0.0125 k_u$, where $k_u \in \mathbb{Z} \cup [1,20]$. We believe this is a novel contribution in the application of POT Hawkes models to financial returns, since, to our knowledge, previous literature has only considered a single (arbitrary) threshold level per study, typically within the range, $0.025 \leq a_u \leq 0.1$. 

Multiple specifications of the 2T-POT Hawkes model were calibrated across this range of threshold levels using the estimation procedure described in \cref{sec:model_select}. Likelihood ratio tests were used to select which variant of the model would be used in the later forecasting validation: this selected for the common intensity model with the $a_{\lambda}^{\rsTA} = 2 a_u \rsUdt^{-1}$ constraint applied and for a Student-$t$ distributed bulk. Tables containing the results of these likelihood ratio tests are included in \cref{sec:model_select} along with a graphical presentation of the estimated parameters (with standard errors) of the selected $H_{2}^{\rsDistT}{(a_u)}$ model as a function of $a_u$. The fit of the exceedance model to both the arrivals process and the distribution of excess magnitudes is tested through residual analysis \citep{Tomlinson2021}. For the arrivals process, the Kolmogorov-Smirniov (KS) test is used to test the null hypothesis that the exceedance arrivals are unit Poisson in residual time,
\begin{equation}
\label{eqn:t_resid}
\tilde{t}^{\rsTO} = \int_{0}^{t}{\lambda^{\rsTO}{(t')} dt'} .
\end{equation}
For the excess distributions, the null hypothesis that the residual excess magnitudes
\begin{equation}
\label{eqn:m_resid}
\tilde{m}_{k}^{\rsTO} = \frac{1}{\xi^{\rsTO}} \ln{\left[1 + \xi^{\rsTO} \frac{m_{k}^{\rsTO}}{\sigma_{t_k}^{\rsTO}}\right]} ,
\end{equation}
are unit exponential distributed is subject to the KS test. These tests are shown graphically in \cref{fig:DAX_tick_m1} for the $H_{2}{(a_u=0.025)}$ exceedance model fitted to the DAX daily log-returns. Equivalent plots for the other five indices are included in the supplementary material.

\begin{figure*}
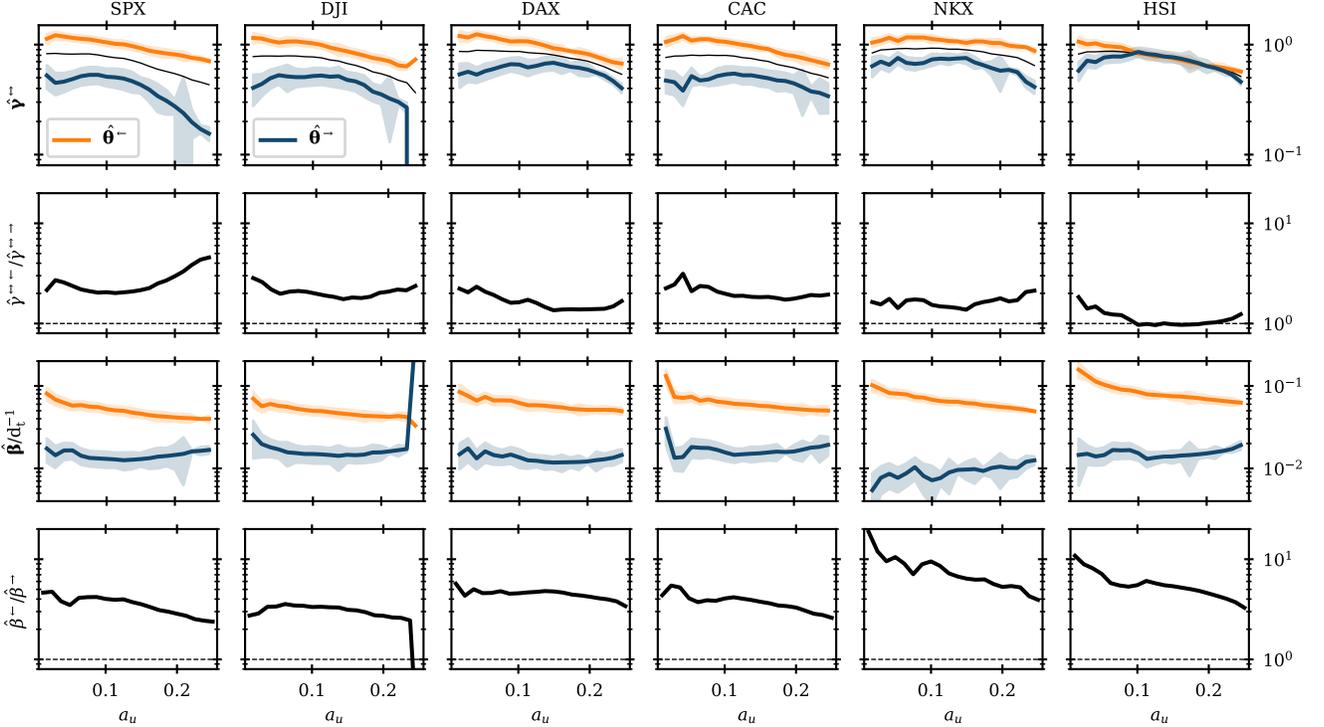

\includegraphics{\rsFolderSlash{\rsTickTrainfolder{ALL}}_SLSQP_gamma_beta_lns_tt_\rsHwkMod_\rsHwkMode_mark}
\caption{
\label{fig:SLSQP_gamma_beta}
Asymmetries in the Hawkes arrival parameters for $H_{2}{(a_u)}$ estimated over the in-sample period, \rsDateTrainStart{} to \rsDateTrainEnd{}. Estimated left- (light orange) and right-tail (dark blue) components of the branching vector $\vect{\upgamma}^{\rsTA}$ (first row, thin black line is the combined branching ratio) and decay constant $\vect{\upbeta}$ (third row) are shown above their respective left-over-right component ratios (second and fourth rows). Vertical axes are displayed on a log-scale.
}
\end{figure*}

A further aim of fitting the Hawkes exceedance models across a wide range of threshold levels $a_u$ is to verify whether the asymmetric Hawkes arrival dynamics reported in \cite{Tomlinson2021} persist across this range. They fitted the 2T-POT Hawkes model to the daily log-returns of the SPX over the in-sample period, 1959-10-02 to 2008-09-01, at $a_{u} = 0.025$ and found significant asymmetries in both the branching vector ($\hat{\gamma}_{\rsTA\rsTL}/\hat{\gamma}_{\rsTA\rsTR} = 2.2 \pm 0.5$) and the decay vector ($\hat{\beta}_{\rsTL}/\hat{\beta}_{\rsTR} = 4.6 \pm 1.2$) that together described a temporal leverage effect. It is anticipated that the parameters will converge to symmetry as $a_u \to 0.5$, since, in this limiting case, every value within $X_t$ will qualify as an exceedance event: thus, the conditional intensity would be fixed and entirely attributed to the exogenous background, making excitation from either tail equally redundant. \cref{fig:SLSQP_gamma_beta} shows that, with some exceptions, the estimated asymmetries in the arrival parameters are remarkably stable along $a_u$ and between the different indices. A slight convergence towards symmetry in both parameters is generally observed as $a_u$ increases, but there is no evidence of a sharp transition within the investigated range. The left-over-right component ratios are quantitatively similar to those reported in \cite{Tomlinson2021} in the vast majority of cases and are never observed to invert. A notable outlier is the HSI branching vector $\hat{\vect{\gamma}}^{\rsTO,\mathrm{HSI}}$, which converges to symmetry at $a_u = 0.1$. This possibly reflects that the HSI had a notably higher average volatility in the in-sample period compared to the other indices (as shown in \cref{tab:data_stat}), which corresponds to the fact that the HSI was a less developed market in this period and, therefore, possibly should be regarded as a separate class. It is also noted that the NKX and HSI decay constant asymmetries $\hat{\beta}^{\rsTL}/\hat{\beta}^{\rsTR}$ are much larger at low $a_u$ compared to the other indices.

Overall, these results add further empirical support to hypothesis of \cite{Tomlinson2021} that there is a temporal aspect to the leverage effect observed in financial daily log-returns, such that the impact of losses on future extremes is not only greater but also more immediate. This important structural feature contributes to the forecasting of exceedance arrivals, which will be seen in the relative forecasting performance of $H_2{(a_u)}$ compared to the fully symmetric $H_1{(a_u)}$ model in \cref{sec:CT_QE}.

\section{\label{sec:CT_QE}Backtesting of out-of-sample quantile forecasts}

In this section, we use the models calibrated on the in-sample data (\rsDateTrainStart{} to \rsDateTrainEnd{}) in \cref{sec:Data} to produce next step ahead forecasts of the left- and right-tail conditional quantile $\hat{\rsVaR}_{a_q, t}^{\rsTO}$ and conditional violation expectation $\hat{\rsES}_{a_q, t}^{\rsTO}$ in the out-of-sample period (\rsDateTrainEnd{} to \rsDateAllEnd{}); the accuracy of these forecasts is then validated and compared through backtesting methods. The forecasts produced by the $H_{2}^{\rsDistT}{(a_u)}$ Hawkes model are compared against those produced by the $G_{1}^{\rsDistT}{(a_u)}$ GARCH-EVT model in order to determine which conditional EVT approach -- Hawkes-type arrival dynamics or GARCH-type volatility dynamics -- provides the best description of extreme log-returns. We also include $H_{1}^{\rsDistT}{(a_u)}$ in this comparison to demonstrate explicitly the benefits of incorporating asymmetry within the 2T-POT Hawkes model. We also take into account three standard GARCH models of increasing complexity [$G_{0}^{\rsDistNorm}{(0)}$, $G_{0}^{\rsDistT}{(0)}$, $G_{1}^{\rsDistT}{(0)}$]. Forecasts are evaluated across the range of coverage levels: $a_q = 0.0025 k_q$, where $k_q \in \mathbb{Z} \cup [1,60]$. This is a much wider and finer range of coverage levels than has been investigated with these backtesting methods in previous financial risk literature \citep{BienBarkowska2020, Taylor2020, Echaust2020, Jalal2008}: this is intended to more fully compare where the relative advantages of each model lie and to explore how this depends on the threshold level $a_u$. Since each of the backtesting methods are defined identically for the left- and the right-tail, the $\rsTO$ superscript is dropped from $\rsVaRI_{a_q, t}^{\rsTO}$, $\rsVaR_{a_q, t}^{\rsTO}$, and $\rsES_{a_q, t}^{\rsTO}$ in the remainder of \cref{sec:CT_QE} for clarity unless explicitly required.

\subsection{\label{ssec:CT_Q}Conditional quantile backtesting methods}

The first set of backtesting methods evaluate the accuracy of the conditional quantile forecasts $\hat{\rsVaR}_{a_q, t}$ by considering the series of observed violations $\hat{\rsVaRI}_{a_q, t}$. If the conditional quantile forecasts are perfectly accurate (i.e. $\hat{\rsVaR}_{a_q, t} = \rsVaR_{a_q, t} \forall t$), then the arrival process of violations is a Poisson point process,
\begin{equation}
\label{eqn:CT_Q_null}
\mathrm{Pr}{\{\hat{\rsVaRI}_{a_q, t}=1 | t\}} = \mathrm{Pr}{\{\rsVaRI_{a_q, t}=1 | t\}} \equiv a_q \forall t .
\end{equation}
All of the backtesting methods in \cref{ssec:CT_Q} derive their null hypothesis from \cref{eqn:CT_Q_null}.

\subsubsection{\label{sssec:CT_Q_UC}Unconditional coverage (UC) test}

The unconditional convergence (UC) test \citep{Kupiec1995} assesses the null hypothesis that the observed proportion of violations $\hat{\pi}_{1}$ is equal to the assumed coverage level $a_q$ ($\mathcal{H}_0 : \hat{\pi}_{1} = a_q$). The UC test is formulated as a likelihood ratio test which compares two Bernoulli likelihood functions. Asymptotically, as the total number of observations in the sample, $T$, goes to infinity, the test statistic $\mathrm{LR}_{\mathrm{UC}}$ is distributed as $\chi^2$ with one degree of freedom:
\begin{equation}
\label{eqn:test_UC_LR}
\mathrm{LR}_{\mathrm{UC}} = -2\ln{\left[
\frac{
a_q^{\hat{T}_1}\left(1-a_q\right)^{1-\hat{T}_1}
}{
\hat{\pi}_{1}^{\hat{T}_1}\left(1 - \hat{\pi}_{1}\right)^{1-\hat{T}_1}}\right]} \sim \chi_1^2 ,
\end{equation}
where $\hat{T}_1 = \sum_t{\hat{\rsVaRI}_{a_q,t}}$ is the observed number of violations in the sample of length $T$ and $\hat{\pi}_{1} = \hat{T}_1/T$. $\mathcal{H}_0$ is rejected if the conditional quantile forecasts $\hat{\rsVaR}_{a_q, t}$ have a significant bias, such that they consistently under- or overestimate the true conditional quantiles $\rsVaR_{a_q, t}$ in the sample.

\subsubsection{\label{sssec:CT_Q_CC}Conditional coverage (CC) test}

The conditional coverage (CC) test \citep{Christoffersen1998} seeks to verify both the correct coverage and the independence of violations over consecutive observations. The process of violations is described by a first-order Markov model, with the CC test based upon the $2 \times 2$ estimated transition matrix with elements $\hat{\pi}_{ij}$ that give the estimated conditional probability of there being a violation ($i=1$) or no violation ($i=0$) at $t$ given that there was ($j=1$) or was not ($j=0$) a violation at $t-1$,
\begin{equation}
\label{eqn:test_CC_matrix}
\hat{\pi}_{ij} = \hat{T}_{ij}/\left(\hat{T}_{i0}+\hat{T}_{i1}\right) ,
\end{equation}
where $\hat{T}_{ij}$ is the number of observations where $\hat{\rsVaRI}_{a_q, t} = i | \hat{\rsVaRI}_{a_q, t-1} = j$. The null hypothesis of the CC test is that the conditional probability of a violation at $t$ with and without a violation at $t-1$ are both equal to the assumed coverage level, i.e.\ENLA{} $\mathcal{H}_0 : \hat{\pi}_{10} = \hat{\pi}_{11} = a_q$. As the total number of observations $T$ goes to infinity, the test statistic $\mathrm{LR}_{\mathrm{CC}}$ is asymptotically distributed as a $\chi^2$ with two degrees of freedom:
\begin{equation}
\label{eqn:test_CC_LR}
\mathrm{LR}_{\mathrm{CC}} = -2\ln{\left[
\frac{
a_q^{\hat{T}_1}\left(1-a_q\right)^{1-\hat{T}_1}
}{
\left(1-\hat{\pi}_{10}\right)^{\hat{T}_{00}}\hat{\pi}_{10}^{\hat{T}_{10}}
\left(1-\hat{\pi}_{11}\right)^{\hat{T}_{01}}\hat{\pi}_{11}^{\hat{T}_{11}}
}\right]} \sim \chi_2^2 .
\end{equation}
$\mathcal{H}_0$ is rejected if either the conditional quantile forecasts at time $t$ that follow a violation at $t-1$ ($\hat{\rsVaR}_{a_q, t} | \hat{\rsVaRI}_{a_q, t-1}=1$) or that follow no violation at $t-1$ ($\hat{\rsVaR}_{a_q, t} | \hat{\rsVaRI}_{a_q, t-1}=0$) have a significant overall bias in the sample.

\subsubsection{\label{sssec:CT_Q_DQ}Dynamic quantile ($DQ_J$) test}

The dynamic quantile ($\mathrm{DQ}_J$) test \citep{Engle2004} is designed to detect higher-order autocorrelation in the series of violations as well as dependence on other explanatory variables. The test is based upon the hit function,
\begin{equation}
\label{eqn:test_DQ_hit}
\hat{\mathrm{Hit}}_{a_q, t} = \hat{\rsVaRI}_{a_q, t} - a_q .
\end{equation}
If follows from \cref{eqn:CT_Q_null,eqn:test_DQ_hit} that, under the correctly specified model, the series $\hat{\mathrm{Hit}}_{a_q, t}$ should be i.i.d.\@ and have zero mean. Accordingly, $\hat{\mathrm{Hit}}_{a_q, t}$ should be independent of lagged values of itself and also independent of the contemporaneous conditional quantile $\hat{\rsVaR}_{a_q, t}$, such that the conditional expectation of $\hat{\mathrm{Hit}}_{a_q, t}$ should be $0$ regardless of any such information available at $t-1$. The $\mathrm{DQ}_J$ test used here is derived as the Wald statistic from an auxiliary regression:
\begin{equation}
\label{eqn:test_DQ_regression}
\hat{\mathrm{Hit}}_{a_q, t} = \phi_0 + \sum_{j=1}^{J}{\phi_{j} \hat{\mathrm{Hit}}_{t-j}} + \phi_{1+J} \hat{\rsVaR}_{a_q, t} + \epsilon_t ,
\end{equation}
where the $\mathrm{DQ}_J$ null hypothesis is $\mathcal{H}_0 : \phi_k = 0 \:\: \forall k \in \mathbb{Z}_0 \cup [0,1+J]$. In other words, the null states that the observed violation coverage probability is equal to the assumed coverage level ($\phi_0 = 0$) and that there is no dependence of $\hat{\mathrm{Hit}}_{a_q, t}$ on the $1+J$ explanatory variables. The DQ test statistic is asymptotically $\chi^2$ distributed with $2+J$ degrees of freedom:
\begin{equation}
\label{eqn:test_DQ_LR}
\mathrm{DQ}_J = \frac{
\hat{\vect{Hit}}_{a_q}'\vect{A}\left(\vect{A}'\vect{A}\right)^{-1}\vect{A}'\hat{\vect{Hit}}_{a_q}
}{
a_q\left(1 - a_q\right)
} \sim \chi_{2+J}^2 ,
\end{equation}
where $\hat{\vect{Hit}}_{a_q}$ is a $1 \times T$ vector containing the series $\hat{\mathrm{Hit}}_{a_q, t}$ observed within the sample and $\vect{A}$ is a $T \times (2 + J)$ matrix comprised of the observed sample series of each explanatory variable.

\subsection{\label{ssec:CT_E}Conditional violation expectation backtesting methods}

\subsubsection{\label{sssec:CT_E_ZMD}Zero mean discrepancy (ZMD) test}

\begin{table*}
\caption{\label{tab:r_CT_agg}
Summari\ENz{}ed backtesting results in the out-of-sample period, \rsDateTrainEnd{} to \rsDateAllEnd{}. The proportions of null hypothesis rejections at the 95\% confidence level ($p<0.05$) across all six indices within coverage level bands $a_q^0 < a_q \leq a_q^1$ are shown for each model, with the results for the EVT models aggregated over three representative values of the threshold level, $a_u \in \left\{0.05,0.1,0.2\right\}$. Lower rejection proportions correspond to better performance on the test; intra-tail performance rankings along each row are given in parentheses, with the first- and second-best performing models highlighted in dark and light gr\ENe{}y, respectively.}
\FTABrCTagg
\end{table*}

\begin{table*}
\caption{\label{tab:r_CT_evt}
Summari\ENz{}ed backtesting results for the EVT models in the out-of-sample period, \rsDateTrainEnd{} to \rsDateAllEnd{}. For each of the EVT models at three values of the threshold level, $a_u \in \left\{0.05,0.1,0.2\right\}$, the proportions of null hypothesis rejections at the 95\% confidence level ($p<0.05$) across all six indices within coverage level bands $a_q^0 < a_q \leq a_q^1$ are shown. Lower rejection proportions correspond to better performance on the test; intra-tail performance rankings along each row are given in parentheses, with the first- and second-best performing models highlighted in dark and light gr\ENe{}y, respectively.}
\FTABrCTevt
\end{table*}

The conditional violation expectation forecasts $\hat{\rsES}_{a_q,t}$ are evaluated by testing the null hypothesis that the standardi\ENz{}ed discrepancies
\begin{equation}
\label{eqn:stand_discrep}
\hat{\rsESD}_{a_q, t} = \frac{X_t - \hat{\rsES}_{a_q, t}}{\hat{\rsVaR}_{a_q, t} - \hat{\rsVaR}_{0.5, t}} ,
\end{equation}
at violations (i.e.\ENLA{} when $\hat{\rsVaRI}_{a_q, t}=1$) have zero mean \citep{McNeil2000}, i.e.\ENLA{} $\mathcal{H}_0: \sum_{t}{\hat{\rsESD}_{a_q, t} \hat{\rsVaRI}_{a_q, t}} / \hat{T}_1 = 0$. Assumptions about the distributions of the standardi\ENz{}ed discrepancies are avoided by employing the dependent circular block bootstrap used in \cite{Jalal2008, Politis2004}. The null hypothesis is subject to a two-tailed test against the alternative hypothesis, $\mathcal{H}_1: \sum_{t}{\hat{\rsESD}_{a_q, t} \hat{\rsVaRI}_{a_q, t}}/\hat{T}_1 =  0$. Thus, the null is rejected when there is a significant bias in the conditional violation expectation forecasts $\hat{\rsES}_{a_q, t}$ over the observed violations ($\hat{\rsVaRI}_{a_q, t}=1$) within the sample.

\subsection{\label{ssec:CT_res}Backtesting results and discussion}

\begin{figure*}
\includegraphics{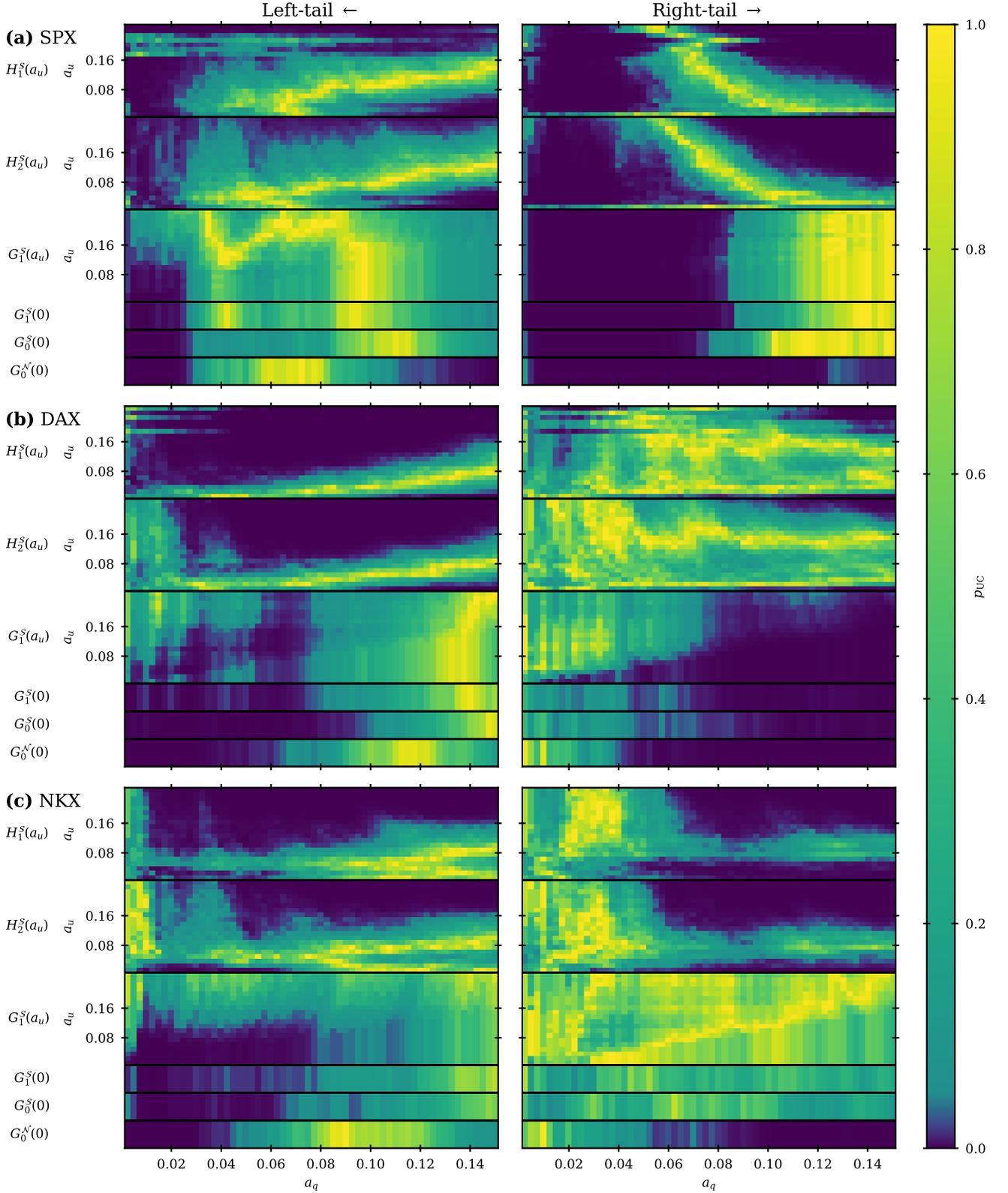}
\caption{
\label{fig:CT_Q_UC_m1_d1_p_1}
Unconditional convergence test $p$-values $p_{\mathrm{UC}}$ at coverage level $a_q$ for three of the six indices (see \cref{fig:CT_Q_UC_m1_d1_p_2} for the other three) in the out-of-sample period, \rsDateTrainEnd{} to \rsDateAllEnd{}. The left- and right-column panels show the results for the left and right tails, respectively, with each panel row giving the results for a particular index: (a) SPX, (b) DAX, and (c) NKX. Within each panel, the $p$-values under six different models (leftmost labels) are shown: the top three (thick) strips show the $p$-values under the EVT models [$H_{1}^{\rsDistT}{(a_u)}$, $H_{2}^{\rsDistT}{(a_u)}$, and $G_{1}^{\rsDistT}{(a_u)}$] as a function of $(a_q, a_u)$; the bottom three (thin) strips show the $p$-values under the standard GARCH models [$G_{1}^{\rsDistT}{(0)}$, $G_{0}^{\rsDistT}{(0)}$, and $G_{0}^{\rsDistNorm}{(0)}$] as a function of $a_q$.
}
\end{figure*}

\begin{figure*}
\includegraphics{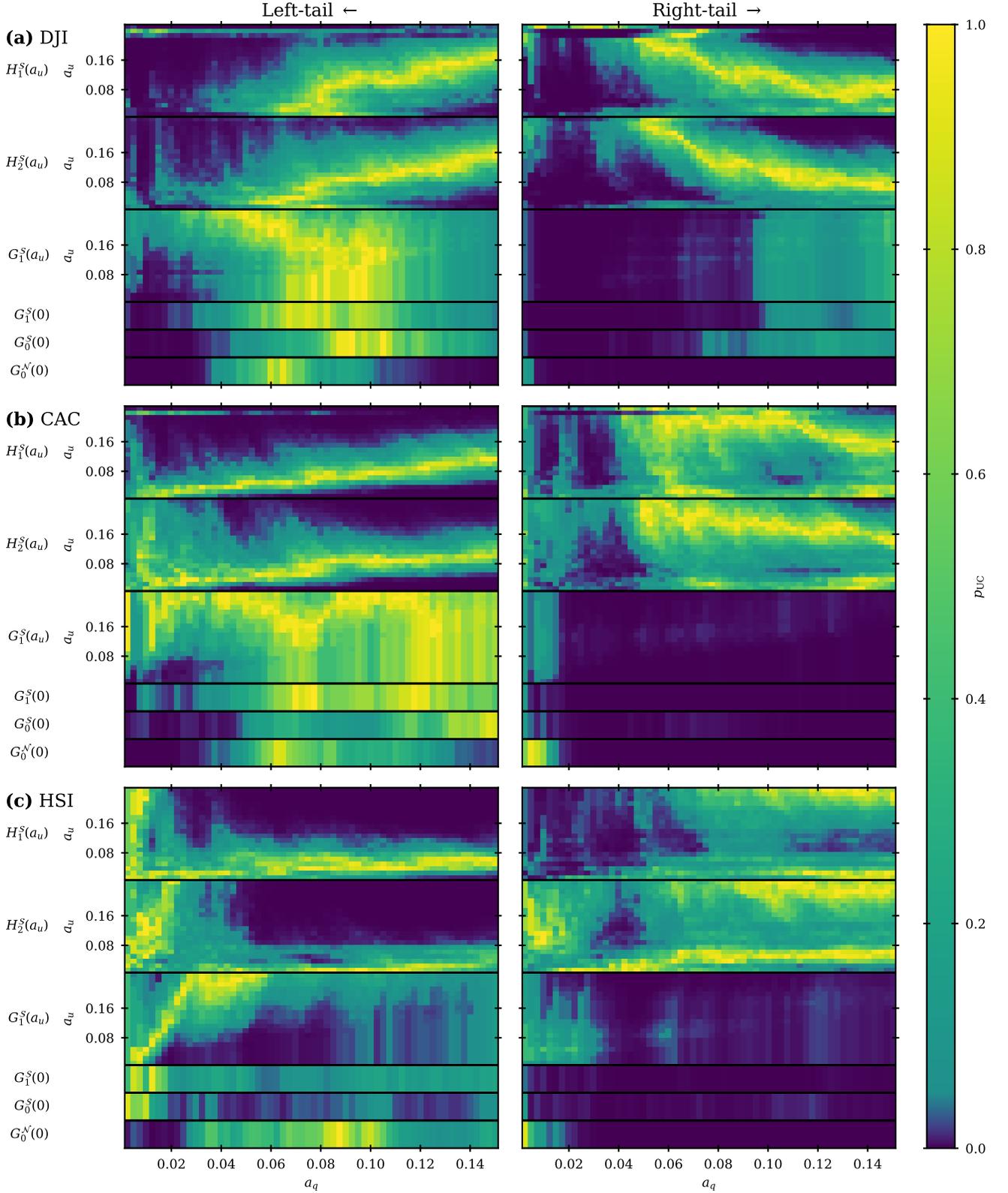}
\caption{
\label{fig:CT_Q_UC_m1_d1_p_2}
Unconditional convergence test $p$-values $p_{\mathrm{UC}}$ at coverage level $a_q$ for three of the six indices (see \cref{fig:CT_Q_UC_m1_d1_p_1} for the other three) in the out-of-sample period, \rsDateTrainEnd{} to \rsDateAllEnd{}. The left- and right-column panels show the results for the left and right tails, respectively, with each panel row giving the results for a particular index: (a) DJI, (b) CAC, and (c) HSI. Within each panel, the $p$-values under six different models (leftmost labels) are shown: the top three (thick) strips show the $p$-values under the EVT models [$H_{1}^{\rsDistT}{(a_u)}$, $H_{2}^{\rsDistT}{(a_u)}$, and $G_{1}^{\rsDistT}{(a_u)}$] as a function of $(a_q, a_u)$; the bottom three (thin) strips show the $p$-values under the standard GARCH models [$G_{1}^{\rsDistT}{(0)}$, $G_{0}^{\rsDistT}{(0)}$, and $G_{0}^{\rsDistNorm}{(0)}$] as a function of $a_q$.
}
\end{figure*}

\begin{figure*}
\includegraphics{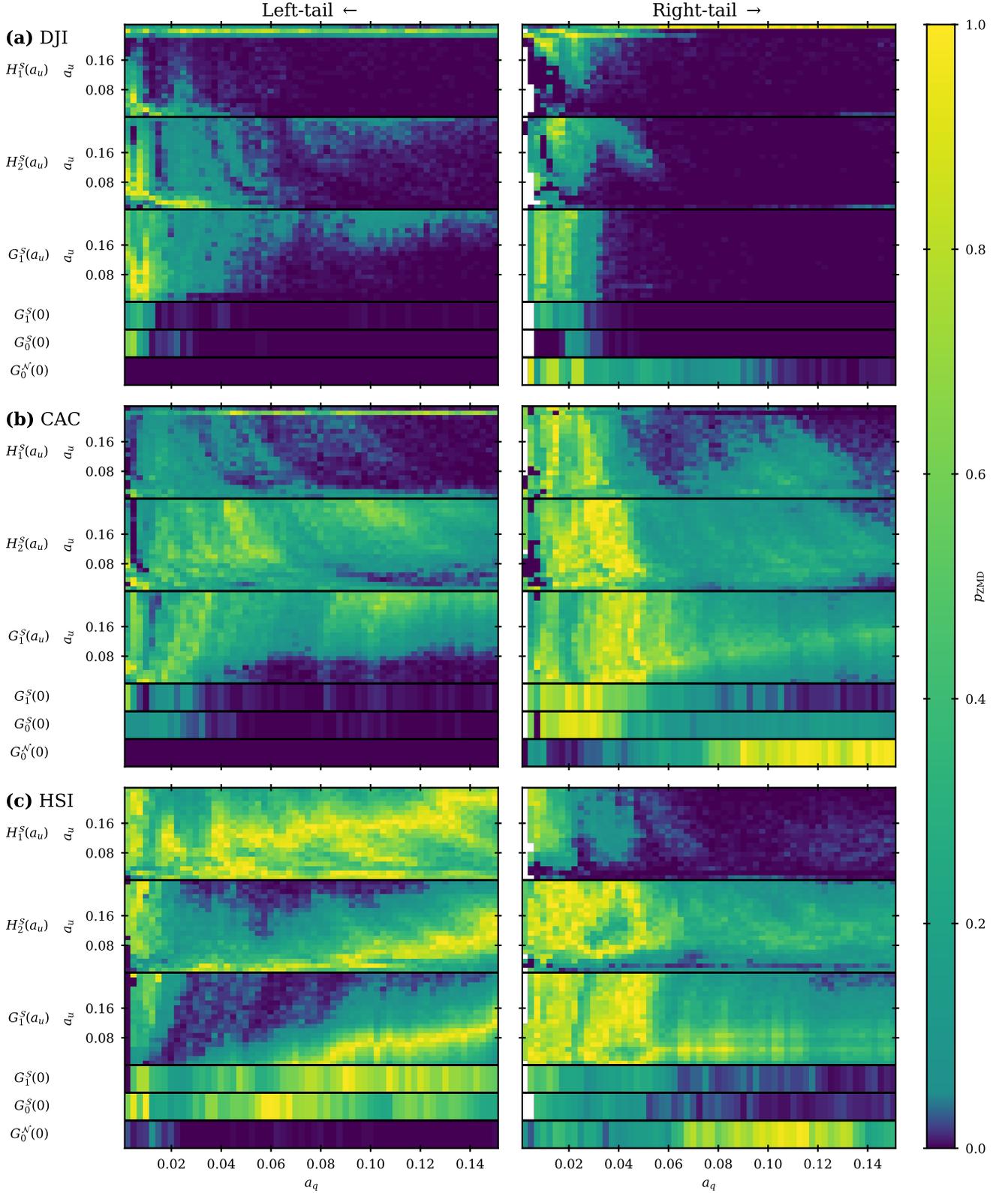}
\caption{
\label{fig:CT_E_ZMD_m1_d1_pLR}
Zero mean discrepancy test $p$-values $p_{\mathrm{ZMD}}$ at coverage level $a_q$ for three of the six indices (see the supplementary material for the other three) in the out-of-sample period, \rsDateTrainEnd{} to \rsDateAllEnd{}. The left- and right-column panels show the results for the left and right tails, respectively, with each panel row giving the results for a particular index: (a) DJI, (b) CAC, and (c) HSI. Within each panel, the $p$-values under six different models (leftmost labels) are shown: the top three (thick) strips show the $p$-values under the EVT models [$H_{1}^{\rsDistT}{(a_u)}$, $H_{2}^{\rsDistT}{(a_u)}$, and $G_{1}^{\rsDistT}{(a_u)}$] as a function of $(a_q, a_u)$; the bottom three (thin) strips show the $p$-values under the standard GARCH models [$G_{1}^{\rsDistT}{(0)}$, $G_{0}^{\rsDistT}{(0)}$, and $G_{0}^{\rsDistNorm}{(0)}$] as a function of $a_q$.
}
\end{figure*}

We present the backtesting results synoptically in \cref{tab:r_CT_agg,tab:r_CT_evt} and as full visuali\ENz{}ations in  \cref{fig:CT_Q_UC_m1_d1_p_1,fig:CT_Q_UC_m1_d1_p_2,fig:CT_E_ZMD_m1_d1_pLR}. \cref{tab:r_CT_agg,tab:r_CT_evt} summari\ENz{}e the results for each backtesting method by giving the proportion of null hypothesis rejections at the 95\% confidence level ($p<0.05$, which appears as dark blue in \cref{fig:CT_Q_UC_m1_d1_p_1,fig:CT_Q_UC_m1_d1_p_2,fig:CT_E_ZMD_m1_d1_pLR}) aggregated across all six indices within six bands of the coverage level, $a_q^0 < a_q \leq a_q^1$. For the EVT models ($a_u>0$), these tables include the results at three representative values of the threshold level, $a_u \in \left\{0.05,0.1,0.2\right\}$. In \cref{tab:r_CT_agg}, the results at these three threshold levels are aggregated together to give a single overall proportion of null rejections for each of the EVT models; these are given with the equivalent results for the standard ($a_u=0$) GARCH models. In \cref{tab:r_CT_evt}, the results for the EVT models at the three representative threshold levels are given independently. These tables are designed so that the relative performance of the models at each test and coverage level can be more easily compared; we are primarily focused on the performance within the lower (i.e.\ENLA{} more extreme) coverage levels, $a_q \leq 0.05$. In addition to these tables, we provide full visualisations of the original $p$-values as a field within the $(a_q,a_u)$-space. Due to the large size of these figures, we only include three within the main paper: \cref{fig:CT_Q_UC_m1_d1_p_1,fig:CT_Q_UC_m1_d1_p_2} shows the full results of the UC test for each of six indices, while \cref{fig:CT_E_ZMD_m1_d1_pLR} shows the results of the ZMD test for the DJI, CAC, and HSI. The equivalent visuali\ENz{}ations for the other tests and data series are included in the supplementary material.

We first observe that the asymmetric $H_{2}^{\rsDistT}{(a_u)}$ model consistently produces more accurate forecasts -- as measured by lower proportions of null rejections -- than the symmetric $H_{1}^{\rsDistT}{(a_u)}$. This is true for both the  aggregated results in \cref{tab:r_CT_agg} and when the two models are compared at the same threshold level $a_u$ in \cref{tab:r_CT_evt}. In the left-tail, this advantage is mostly restricted to the lower coverage bands; specifically, to $a_{q} \leq 0.05$ in the UC and $\mathrm{DQ}_{4}$ tests, to $a_{q} \leq 0.075$ in the CC test, and to the full range, $a_{q} \leq 0.15$, in the ZMD test. In the right-tail, $H_{2}^{\rsDistT}{(a_u)}$ tends to hold an advantage over $H_{1}^{\rsDistT}{(a_u)}$ at all values of $a_q$ in the UC, CC, and ZMD tests, but this advantage is larger in lower coverage bands and there is no consistent trend for the $\mathrm{DQ}_{4}$ test. It can be taken from this that there is a demonstrable gain in predictive power when asymmetries are incorporated within the 2T-POT Hawkes model, and this gain is strongest for the most extreme future events. This adds further empirical evidence for the role of these asymmetries -- including those that describe the temporal leverage effect -- in the data generating process for extreme daily log-returns in high-cap equity indices.

Now considering the GARCH-type models in \cref{tab:r_CT_agg}, it is observed for the left-tail that increasing model complexity does tend to yield more accurate forecasts, especially within the lowest coverage level band, $0 < a_{q} \leq 0.025$. This shows that each of the additional features -- the leptokurtic innovation distribution [$G_{0}^{\rsDistT}{(0)}$], the leverage effect in the variance dynamics [$G_{1}^{\rsDistT}{(0)}$], and the asymmetric GP tails [$G_{1}^{\rsDistT}{(a_u)}$] -- capture significant aspects of the generating process for left-tail extremes. However, for the right-tail extremes, the simplest $G_{0}^{\rsDistNorm}{(0)}$ model outperforms the two more complex standard ($a_u=0$) GARCH models in all of the tests based on the conditional quantile (UC, CC, and $\mathrm{DQ}_4$) within this lowest coverage band. This is because the Student-$t$ distribution used for the innovations $\epsilon_t$ in $G_{0}^{\rsDistT}{(0)}$ and $G_{1}^{\rsDistT}{(0)}$ overestimates the heaviness of the right-tail and so provides a worse approximation than the normal distribution. This problem is to be expected when a symmetric innovation distribution is fitted to the full distribution of observations in one step, since the fit to the right-tail is compromised by the significantly heavier left-tail. In contrast, the asymmetric GP tails of the GARCH-EVT model are fitted to each tail independently, and so the negative impact on the forecasting accuracy of $G_{1}^{\rsDistT}{(a_u)}$ in both tails is mitigated, especially at the best performing threshold level in \cref{tab:r_CT_evt}. This independence demonstrates a key advantage of conditional EVT methods to forecasters of extreme events.

We now compare the asymmetric 2T-POT Hawkes model [$H_{2}^{\rsDistT}{(a_u)}$] with the GARCH-EVT model [$G_{1}^{\rsDistT}{(a_u)}$], focusing first on the results for the UC and CC tests. We reiterate that these are simple convergence tests that verify whether the fraction of conditional quantile violations in the out-of-sample period is equal to the coverage level $a_q$ (UC test) and also if the series $\hat{\rsVaRI}_{a_q,t}$ is serial independent over a single time step (CC test). These tests therefore provide the most basic measures of overall accuracy and serial independence for the next step ahead forecasts of the conditional quantile $\hat{\rsVaR}_{a_q,t}$. For the threshold aggregated UC and CC test results in \cref{tab:r_CT_agg}, $H_{2}^{\rsDistT}{(a_u)}$ is by far the best performing model within the lowest left-tail coverage band ($a_q \leq 0.025$) and in all right-tail coverage bands ($a_q \leq 0.15$). Looking at the equivalent threshold disaggregated results in \cref{tab:r_CT_evt}, we see that, for the left-tail, $G_{1}^{\rsDistT}{(0.2)}$ and $H_{2}^{\rsDistT}{(0.05)}$ stand out as the first- and second-best performing models within the lower half of coverage bands, $a_q \leq 0.075$. Over the same range of coverage bands in the right-tail, $H_{2}^{\rsDistT}{(a_u)}$ consistently outperforms $G_{1}^{\rsDistT}{(a_u)}$  irrespective of threshold selection, though $H_{2}^{\rsDistT}{(0.2)}$ is by far the best performing. 

To better understand the summari\ENz{}ed UC and CC test results in \cref{tab:r_CT_agg,tab:r_CT_evt}, we examine the visuali\ENz{}ation of the full results of the UC test in \cref{fig:CT_Q_UC_m1_d1_p_1,fig:CT_Q_UC_m1_d1_p_2}. The equivalent visuali\ENz{}ations for the CC test (which can be found in the supplementary material) display very similar features, meaning that the following observations are robust with respect to lag-1 violation independence. A consistent tuning relationship is observed in the left-tail for the 2T-POT Hawkes models across the six indices, wherein there is an approximately linear relationship between coverage level $a_q$ and the optimal threshold level $a_u$ for forecasting accuracy at that coverage level, as observed by the highest average $p$-values and, accordingly, by the lowest density of $p<0.05$ null hypothesis rejections within the $(a_{q}, a_{u})$-space. For the DAX, NKX, and HSI, this trend is flatter, such that very accurate forecasts are produced across almost the full range of coverage levels $a_q$ at a single (low) value of the threshold level $a_u$. This is reflected in \cref{tab:r_CT_evt}, where $H_{2}^{\rsDistT}{(0.05)}$ is the best performing Hawkes model up to $a_{q} \leq 0.125$. This predictable and stable dependency of forecasting accuracy on threshold selection is ideal for practical applications, since it means that the correctly tuned 2T-POT Hawkes model should produce reliably accurate forecasts of the left-tail conditional quantile $\hat{\rsVaR}_{a_q,t}^{\rsTL}$ across a broad range of coverage levels $a_q$. In contrast, the pattern of forecasting accuracy for $G_{1}^{\rsDistT}{(a_u)}$ in the left-tail is much more unpredictable. Phases of poor forecasting accuracy -- as identified by rejection of the null hypothesis -- show erratic dependence on threshold selection and instead tend to concentrate within ranges of coverage levels $a_q$ that are different for each of the indices. Once consistent feature (observed in all but the HSI data) is that $G_{1}^{\rsDistT}{(a_u)}$ produces a phase of poor forecasting in the left-tail when both $a_q$ and $a_u$ are at their lowest. Consequently, the forecasting accuracy of the GARCH-EVT model with respect to the left-tail conditional quantile $\hat{\rsVaR}_{a_q,t}^{\rsTL}$ (i.e.\ENLA{} value-at-risk) at the most extreme coverage levels ($a_q \leq 0.025$) is much more sensitive on threshold selection than that of the 2T-POT Hawkes model; hence it is the latter that has the lowest proportion of null hypothesis rejections in the corresponding rows of \cref{tab:r_CT_agg}. From this, we infer that the 2T-POT Hawkes model is more reliably accurate than the GARCH-EVT model in this most crucial case.

In the right-tail, the tuning relationship for the 2T-POT Hawkes models is found to be inverted, such that the most accurate forecasts for high coverage levels are achieved when the threshold level $a_u$ is low and vice versa. Considering the asymmetries in the arrival parameters reported in \cref{ssec:Estimation}, this likely reflects that less extreme ($a_q > 0.05$) right-tail observations are frequently triggered by the preceding arrivals of more extreme ($a_u<0.05$) left-tail observations. Despite this unexpected pattern, the right-tail forecasting accuracy of $H_{2}^{\rsDistT}{(a_u)}$ retains some of the practical advantages compared to $G_{1}^{\rsDistT}{(a_u)}$ as observed in the left-tail, namely greater consistency between the different indices and approximate tuning relationships that allow for maximally effective forecasting over a wider range of coverage levels $a_q$. Indeed, it is noted that the aggregated right-tail performance of the GARCH-based models in the UC and CC tests in \cref{tab:r_CT_agg,tab:r_CT_evt} would be significantly worse across all coverage bands but for the outlier that is the NKX data. 

Next, we consider the results for the $\mathrm{DQ}_4$ test. As defined in \cref{sssec:CT_Q_DQ}, this test verifies whether the intensity of conditional quantile violations (and, therefore, the accuracy of conditional quantile forecasts $\hat{\rsVaR}_{a_q,t}$) are dependent on a set of explanatory series, including lagged values of $\hat{\rsVaR}_{a_q,t}$ and $\hat{\rsVaRI}_{a_q,t}$. It is expected that the $\mathrm{DQ}_4$ test might be less fav\ENou{}rable to the 2T-POT Hawkes model due to its inherently less smooth response to extreme log-returns. Violations at time $t-1$ ($\hat{\rsVaRI}_{a_q, t-1}=1$) are more likely to also be exceedance events within $N^{\rsTA}$ than are non\ENhyphNon{}violations ($\hat{\rsVaRI}_{a_q, t-1}=0$). This would tend to cause larger increases in the forecast conditional quantile $\hat{\rsVaR}_{a_q,t}$ at time $t$ due to the jump-like response of the Hawkes model to exceedances, which might then cause a negative bias in the coefficient $\phi_1$ in \cref{eqn:test_DQ_regression}. In the threshold aggregated $\mathrm{DQ}_4$ test results (\cref{tab:r_CT_agg}), we find in both tails that $H_{2}^{\rsDistT}{(a_u)}$ and $G_{1}^{\rsDistT}{(a_u)}$ share near equal proportions of null hypothesis rejections in the two lowest coverage bands ($a_q \leq 0.05$), with those for the Hawkes model being slightly lower in the left-tail and slightly higher in the right-tail. For the higher coverage bands ($0.05 < a_q$), $H_{2}^{\rsDistT}{(a_u)}$ begins to perform notably worse on this test relative to $G_{1}^{\rsDistT}{(a_u)}$. In the threshold disaggregated $\mathrm{DQ}_4$ test results (\cref{tab:r_CT_evt}), we again observe the performance of $G_{1}^{\rsDistT}{(a_u)}$ in the extreme left-tail is much more sensitive on threshold selection compared with $H_{2}^{\rsDistT}{(a_u)}$. The main implication of these results is that, at the 5\% coverage level or lower, the left- and right-tail conditional quantile forecasts $\hat{\rsVaR}_{a_q,t}^{\rsTO}$ produced by the 2T-POT Hawkes model show no greater conditional bias than the forecasts produced by the GARCH-EVT model. This is an important point for practical forecasting applications, where such conditional biases might be correlated with risk, and so might inflict a disproportionate penalty. When taken with the UC and CC test results, this shows that the 2T-POT Hawkes model presents a superior alternative to GARCH-EVT for these forecasts.

Finally, we examine the results for the ZMD test, which are summari\ENz{}ed in \cref{tab:r_CT_agg,tab:r_CT_evt} and are visuali\ENz{}ed in \cref{fig:CT_E_ZMD_m1_d1_pLR}. The ZMD test is the only test that measures the accuracy of the conditional violation expectation forecasts $\hat{\rsES}_{a_q,t}$; consequently, it places more emphasis on the accuracy of the tail distribution -- especially at the extremities -- compared to the other tests. This is evident when comparing $H_{2}^{\rsDistT}{(a_u)}$ to $H_{1}^{\rsDistT}{(a_u)}$: the proportions of null hypothesis rejections under the former are universally lower in both tails because of the significant asymmetries found in the shape parameter $\hat{\vect{\upxi}}$ of the GP tail distributions (see \cref{fig:SLSQP} in \cref{sec:model_select}). In comparison to the UC test visuali\ENz{}ed in \cref{fig:CT_Q_UC_m1_d1_p_1,fig:CT_Q_UC_m1_d1_p_2}, the ZMD test results visuali\ENz{}ed in \cref{fig:CT_E_ZMD_m1_d1_pLR} show much weaker sensitivity on the threshold level $a_u$ in all cases. It is also apparent that the test is affected by low sample size at the smallest values of $a_u$. Indeed, there is sometimes no defined value for the test statistic at the lowest values of $a_q$ because there are not enough violations at this coverage level within the out-of-sample period to perform the circular block bootstrap. Notably, this is only observed in the right-tail -- indicating an asymmetric bias in all models that results in a systemic overestimation of the more extreme ($a_q \leq 0.05$) right-tail conditional quantiles $\hat{\rsVaR}_{a_q,t}^{\rsTR}$ (and perhaps also a systemic underestimation of $\hat{\rsVaR}_{a_q,t}^{\rsTL}$ within this range as well).

For the threshold aggregated left-tail results in \cref{tab:r_CT_agg}, ZMD is the test in which the performance of the 2T-POT Hawkes model is strongest relative to the GARCH-EVT model. The two models perform almost equally well in the lowest band ($a_q \leq 0.025$), the $H_{2}^{\rsDistT}{(a_u)}$ is then the best performing model by far in the other coverage bands. Conversely, in the right-tail, $G_{1}^{\rsDistT}{(a_u)}$ is found to have slightly lower proportions of null rejections than $H_{2}^{\rsDistT}{(a_u)}$ in 5 out of 6 coverage bands; this is the opposite of what was observed in the UC and CC tests, where the latter held a clear advantage. One explanation is that tests are technically performed over different samples: the UC and CC tests (which examine the conditional quantile forecasts $\hat{\rsVaR}_{a_q,t}$) consider all observations in $X_t$, whereas the ZMD test only considers observations that are also violations ($\hat{\rsVaRI}_{a_q,t} = 1$). In particular, we note that the right-tail results in the lowest coverage band ($a_q \leq 0.025$) are affected by the failure of the circular block bootstrap observed in \cref{fig:CT_E_ZMD_m1_d1_pLR}. Nevertheless, these results show that the 2T-POT Hawkes model presents a very competitive forecasting tool for financial risk analysts to consider.

\section{\label{sec:Summary}Summary and concluding remarks}

We have developed the 2T-POT Hawkes model of \cite{Tomlinson2021} as a conditional EVT model adapted for the forecasting of conditional quantile-based risk measures in both the left and right tails of the same real univariate discrete time series $X_t$. Our reparameteri\ENz{}ation of the exceedance model in terms of the expected average intensity $a_{\lambda}$ more than halved the optimi\ENz{}ation time, and, when constrained by the threshold level $a_u$, achieved a dimension reduction of \ENone{} at a negligible cost to the goodness of fit, as verified by in- and out-of-sample likelihood ratio tests. The resulting improvements to the speed and reliability of the optimi\ENz{}ation procedure enabled us to greatly expand our application of the 2T-POT Hawkes model to historic market data: (i) the model was fitted to the daily log-returns of six international large cap equity indices over the in-sample period, \rsDateTrainStart{} to \rsDateTrainEnd{}; (ii) we independently fitted the model to each series using a wide range of exceedance thresholds, from the 1.25\% to 25.00\% mirrored in-sample quantiles. The significant asymmetries in the estimated branching vector $\hat{\vect{\upgamma}}^{\rsTA}$ and the decay constant $\hat{\vect{\upbeta}}$ reported in \cite{Tomlinson2021} were reproduced here and were quantitatively similar across the six indices; moreover, these asymmetries were found to be stable over the majority of the tested threshold levels $a_u$. This adds further empirical support for a temporal leverage effect in which the impact of losses is not only greater but also more immediate.

By introducing a subordinate bulk distribution that is conditional upon the Hawkes exceedance process, we extended the support of the 2T-POT Hawkes model to the full distribution of $X_t$; this guaranteed that forecasts of conditional quantile-based risk measures were always defined at all coverage levels, $a_q \in [0,1]$. The fully supported 2T-POT Hawkes model was used to produce next step ahead forecasts of the left- and right-tail conditional quantiles $\hat{\rsVaR}_{a_q,t}^{\rsTO}$ (value-at-risk) and the conditional violation expectations $\hat{\rsES}_{a_q,t}^{\rsTO}$ (expected shortfall). The accuracy and serial independence of these forecasts in the out-of-sample period, \rsDateTrainEnd{} to \rsDateAllEnd{}, were assessed through backtesting methods; these results were compared with those for the symmetric 2T-POT Hawkes model and for a set of GARCH models that included GARCH-EVT. This greatly expanded upon similar analysis in previous literature \citep{BienBarkowska2020, Taylor2020, Echaust2020, Jalal2008}: both by extending the analysis to the right-tail and by evaluating forecasts over a much wider and finer range of coverage levels, from 0.25\% to 15.00\%. Our asymmetric 2T-POT Hawkes model [$H_{2}^{\rsDistT}{(a_u)}$] was found to produce the most reliably accurate conditional quantile forecasts $\hat{\rsVaR}_{a_q,t}^{\rsTO}$ in both tails at the 5\% coverage level or lower. Within this same coverage range, our model was also found to produce the most accurate forecasts of the left-tail conditional violation expectation $\hat{\rsES}_{a_q,t}^{\rsTL}$, though the right-tail forecasts $\hat{\rsES}_{a_q,t}^{\rsTR}$ were slightly less accurate than those produced by the GARCH-EVT model [$H_{1}^{\rsDistT}{(a_u)}$]. The comparison with the symmetric 2T-POT Hawkes model [$H_{1}^{\rsDistT}{(a_u)}$] confirms that the incorporation of left-right asymmetries provides a demonstrable increase in predictive power, which adds further empirical support to the significance of the temporal leverage effect. The comparison with the GARCH-EVT model indicates that asymmetric Hawkes-type arrival dynamics provide a better approximation of the true data generating process for extreme log-returns within large cap equity indices than GARCH-type variance dynamics. This successfully extends the foundational principle of extreme value theory -- that `extreme events should speak for themselves' -- to the conditional case.

In future work, the comparison between Hawkes- and GARCH-based EVT methods could be extended to include time inhomogeneous processes. These could be based, for example, on the MSGARCH process developed in \cite{Ardia2018}, which describes a single GARCH process that stochastically switches between multiple regimes according to a Markov transition matrix. If developed, a regime switching variant of the 2T-POT Hawkes model could identify whether the asymmetries between the tails reported here are constant or change under different market states. In order to extend the comparison of conditional EVT methods done in this paper to the like-for-like time inhomogeneous case, a novel merger of the MSGARCH model of \cite{Ardia2018} and GARCH-EVT model of \cite{McNeil2000} could be developed. Other possible future developments of this work include the extension of the analysis to multi-step ahead aggregate forecasts of the same quantile-based risk measures. This is a natural application for the 2T-POT Hawkes model, since accurate single-step forecasts for one tail would have a compounding impact on the multi-step aggregate forecasts of both tails. To further improve forecasting accuracy, the 2T-POT Hawkes model could also incorporate explanatory variables as sources of non\ENhyph{}constant exogenous intensity. Candidates relevant to the equity indices used here include volatility indices such as the CBOE Volatility Index (VIX). Finally, possible generating mechanisms for the universal asymmetries observed across the six indices may be explored by fitting the 2T-POT Hawkes model to the output of relevant heterogeneous agent-based models (hABMs).

\docAcknowledgeOpen
An earlier version of this paper was uploaded to the arXiv preprint server under the title ``2T-POT Hawkes model for dynamic left- and right-tail quantile forecasts of financial returns: out-of-sample validation of self-exciting extremes versus conditional volatility''. We thank the Associate Editor of the International Journal of Forecasting and the two anonymous referees for their feedback on our original submission. Their comments greatly helped us improve the clarity of our arguments and the presentation of our results.
\ifanon
Funding acknowledgements are redacted for double blind peer review.
\else
M.F.T. acknowledges support from EPSRC (UK) Grant No. EP/R513155/1 and CheckRisk LLP.
\fi
\docAcknowledgeClose

\appendix

\section{\label{sec:reparam}2T-POT Hawkes model reparameteri\ENz{}ation}

In the univariate Hawkes process, $\lambda = \mu + \gamma\chi$, each mother event spawns $\gamma$ daughter events on average. If the original (0-th) generation of events is triggered by the exogenous background intensity $\mu$ and every subsequent generation by the endogenous excitement produced by the preceding generation, then the $n$-th generation events will arrive with an average intensity $\gamma^n \mu$. Thus, by summation, the expected average intensity for the whole process is
\begin{equation}
\label{eqn:a_lambda_uni}
a_{\lambda} 
\equiv \mathbb{E}{\left[\lambda\right]} 
= \sum_{n=0}^{\infty}\gamma^n \mu
= \frac{\mu}{1-\gamma} .
\end{equation}
In \cref{eqn:a_lambda_uni}, the infinite sum $\sum_{n=0}^{\infty}\gamma^n = (1-\gamma)^{-1}$ is the average number of events in a full lineage. Equivalently, it may be said that $1-\gamma$ is the proportion of all events that are in the 0-th generation, hence
\begin{equation}
\label{eqn:Hawkes_mu_constraint_uni}
\mu
= \left(1-\gamma\right) a_{\lambda} .
\end{equation}

In the multivariate case, $\vect{\lambda}=\vect{\upmu}+\vect{\Gamma}\vect{\upchi}$, the full lineage of the branching matrix is
\inOneTwoColumn{
\begin{equation}
\label{eqn:lineage_multi}
\sum_{n=0}^{\infty}\vect{\Gamma}^n
= \sum_{n=0}^{\infty}\left(\vect{Q}\vect{D}\vect{Q}^{-1}\right)^n
= \sum_{n=0}^{\infty}{\vect{Q}\vect{D}^n\vect{Q}^{-1}}
= \vect{Q}\left(\vect{I}-\vect{D}\right)^{-1}\vect{Q}^{-1}
= \left(\vect{I}-\vect{\Gamma}\right)^{-1}
,
\end{equation} 
}{
\begin{align}
\sum_{n=0}^{\infty}\vect{\Gamma}^n
&= \sum_{n=0}^{\infty}\left(\vect{Q}\vect{D}\vect{Q}^{-1}\right)^n
= \sum_{n=0}^{\infty}{\vect{Q}\vect{D}^n\vect{Q}^{-1}}
\nonumber\\
&= \vect{Q}\left(\vect{I}-\vect{D}\right)^{-1}\vect{Q}^{-1}
\nonumber\\
&= \left(\vect{I}-\vect{\Gamma}\right)^{-1}
,
\end{align}
}
where $\vect{Q}$ and $\vect{D}$ are the matrices of eigenvectors and eigenvalues of $\vect{\Gamma}$, respectively, and $\vect{I}$ is the identity matrix. By analogy with \cref{eqn:a_lambda_uni},
\begin{equation}
\label{eqn:a_lambda_multi}
\vect{a}_{\lambda}
\equiv \mathbb{E}{\left[\vect{\uplambda}\right]}  
= \sum_{n=0}^{\infty}\vect{\Gamma}^n  \vect{\upmu}
= \left(\vect{I}-\vect{\Gamma}\right)^{-1} \vect{\upmu} ,
\end{equation}
and, therefore,
\begin{equation}
\label{eqn:Hawkes_mu_constraint_multi}
\vect{\upmu} 
= \left(\vect{I}-\vect{\Gamma}\right) \vect{a}_{\lambda} .
\end{equation}

The common intensity model is implemented as a constrained case of the bivariate process,
\begin{equation}
\label{eqn:bivariate_model_long}
\left(\begin{matrix}
			\lambda^{\rsTL} \\
			\lambda^{\rsTR} \\
\end{matrix}\right)
= 		
\left(\begin{matrix}
			\mu^{\rsTL} \\
			\mu^{\rsTR} \\
\end{matrix}\right)	
+
\left(\begin{matrix}
			\gamma^{\rsTL\rsTL} & \gamma^{\rsTL\rsTR}\\
			\gamma^{\rsTR\rsTL} & \gamma^{\rsTR\rsTR}\\
\end{matrix}\right)
\left(\begin{matrix}
			\chi^{\rsTL} \\
			\chi^{\rsTR} \\
\end{matrix}\right) , 
\end{equation}
in which $a_{\lambda}^{\rsTL} = a_{\lambda}^{\rsTR} = a_{\lambda}^{\rsTA}/2$, $\mu^{\rsTL} = \mu^{\rsTR} = \mu^{\rsTA}/2$, and $\gamma^{\rsTL\rsTO} = \gamma^{\rsTR\rsTO} = \gamma^{\rsTA\rsTO}/2$. Substituting these into \cref{eqn:Hawkes_mu_constraint_multi} yields 
\begin{equation}
\label{eqn:Hawkes_mu_constraint_app}
\mu^{\rsTA} = \left[2 - \left(\gamma^{\rsTA\rsTL} + \gamma^{\rsTA\rsTR}\right)\right] a_{\lambda}^{\rsTA} .
\end{equation}

\section{\label{sec:model_select}Model estimation and selection}

\begin{figure*}
\includegraphics{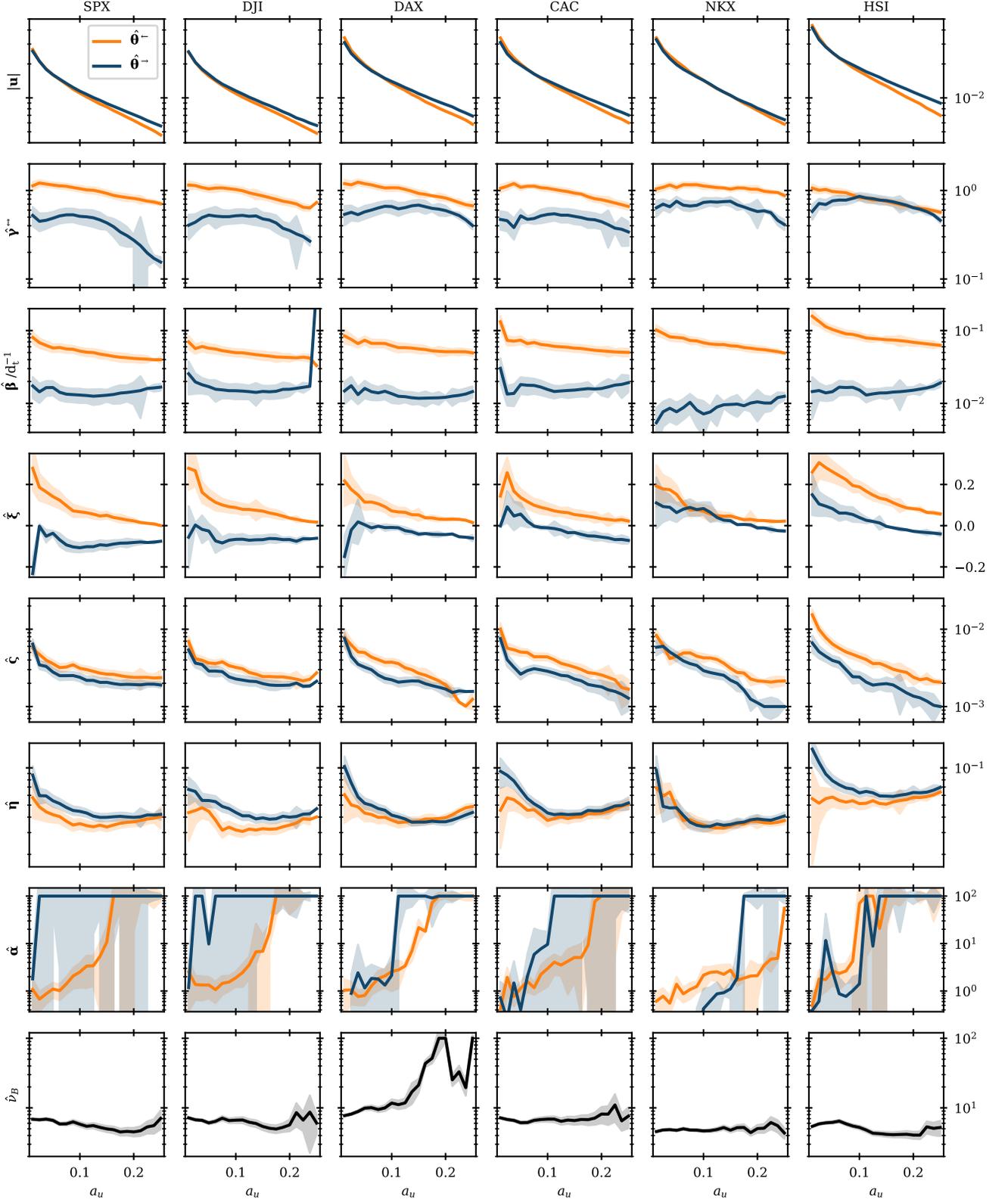}
\caption{
\label{fig:SLSQP}
$H_{2}^{\rsDistT}{(a_u)}$ parameter estimates (lines) and standard errors (shaded areas) calibrated over the in-sample period, \rsDateTrainStart{} to \rsDateTrainEnd{}. The left- and right-tail components of the vector parameters are shown in light orange and dark blue, respectively. Vertical axes (except for those on the row describing the estimated GP shape parameter vector $\hat{\vect{\upxi}}$) are displayed on a log-scale.
}
\end{figure*}

The log-likelihood of the 2T-POT Hawkes exceedance model under the parameters $\vect{\uptheta}_{u}$ over the data $X_{0:T} = \{X_t | t \in \mathbb{Z} \cup [0,T) \}$ can be expressed as the sum over the tails
\begin{equation}
\label{eqn:logL_H_u}
\ell_{u}^{\rsTA}{\left(\vect{\uptheta}_{u} \middle| X_{0:T}\right)}
= \sum_{i \in \{\rsTL, \rsTR\}}{\ell_{u}{^{i}\left(\vect{\uptheta}_{u} \middle| X_{0:T}\right)}
} .
\end{equation}
The log-likelihood components for each tail can be further separated:
\begin{equation}
\label{eqn:logL_H_u_sep}
\ell_{u}^{\rsTO}{\left(\vect{\uptheta}_{u} \middle| X_{0:T}\right)}
= \ell_{\lambda}^{\rsTO}{\left(\vect{\uptheta}_{u} \middle| X_{0:T}\right)}
+\ell_{M}^{\rsTO}{\left(\vect{\uptheta}_{u} \middle| X_{0:T}\right)} ,
\end{equation}
where
\inOneTwoColumn{
\begin{align}
\label{eqn:logL_H_lambda}
\ell_{\lambda}^{\rsTO}{\left(\vect{\uptheta}_{u} \middle| X_{0:T}\right)} 
= &- \int_{0}^{T-1}{\lambda^{\rsTO}{\left(t' \middle|\vect{\uptheta}_{u}; \mathcal{I}_{t'}\right)}d{t'}} 
+ \sum_{k: t_{k}^{\rsTO} < T}{
\ln{\left[\lambda^{\rsTO}{\left(t_{k}^{\rsTO} \middle|\vect{\uptheta}_{u}; \mathcal{I}_{t_{k}^{\rsTO}}\right)}\right]} 
} ,
\end{align}
\begin{equation}
\label{eqn:logL_H_M}
\ell_{M}^{\rsTO}{\left(\vect{\uptheta}_{u} \middle| X_{0:T}\right)} 
= \sum_{k: t_{k}^{\rsTO} < T}{
\ln{\left[f_{M,t_{k}^{\rsTO}}^{\rsDistGP,\rsTO}{\left(M_{k}^{\rsTO} \middle| \vect{\uptheta}_{u}; \mathcal{I}_{t_{k}^{\rsTO}} \right)}\right]}
} ,
\end{align}
}{
\begin{align}
\label{eqn:logL_H_lambda}
\ell_{\lambda}^{\rsTO}{\left(\vect{\uptheta}_{u} \middle| X_{0:T}\right)} 
= &- \int_{0}^{T-1}{\lambda^{\rsTO}{\left(t' \middle|\vect{\uptheta}_{u}; \mathcal{I}_{t'}\right)}d{t'}} 
\nonumber\\
&+ \sum_{k: t_{k}^{\rsTO} < T}{
\ln{\left[\lambda^{\rsTO}{\left(t_{k}^{\rsTO} \middle|\vect{\uptheta}_{u}; \mathcal{I}_{t_{k}^{\rsTO}}\right)}\right]} 
} ,
\end{align}
\begin{equation}
\label{eqn:logL_H_M}
\ell_{M}^{\rsTO}{\left(\vect{\uptheta}_{u} \middle| X_{0:T}\right)} 
= \sum_{k: t_{k}^{\rsTO} < T}{
\ln{\left[f_{M,t_{k}^{\rsTO}}^{\rsDistGP,\rsTO}{\left(M_{k}^{\rsTO} \middle| \vect{\uptheta}_{u}; \mathcal{I}_{t_{k}^{\rsTO}} \right)}\right]}
} ,
\end{equation}
}
are the log-likelihood components of the arrivals process and of the conditional GP tail distributions, respectively. The estimated exceedance model parameters $\hat{\vect{\uptheta}}_{u}$ are then estimated by ML estimation using the SLSQP method in SciPy \citep{Nocedal2006c18}. Any unconstrained parameters of the parametric bulk distribution (denoted by the vector $\vect{\uptheta}_{B}^{\rsDistD}$) are obtained in a second step, through maximi\ENz{}ation of the log-likelihood
\begin{equation}
\label{eqn:logL_H_B}
\ell_{B}^{\rsDistD}{\left(\vect{\uptheta}_{B}^{\rsDistD} \middle| X_{0:T}\right)}
= \sum_{t \notin \{t_{k}^{\rsTA}\} < T}{
\ln{\left[f_{B,t}^{\rsDistD}{\left(X_{t} \middle| \vect{\uptheta}_{B}^{\rsDistD}\right)}\right]}
} .
\end{equation}
\cref{fig:SLSQP} shows the estimated parameters of the $H_{2}^{\rsDistT}{(a_u)}$ model as a function of $a_u$. The standard errors of the estimated parameters are obtained by finite difference approximation of the Hessian matrix.

Alternative parameteri\ENz{}ations of the 2T-POT Hawkes model are compared through likelihood ratio tests in order to select the appropriate parameteri\ENz{}ation for the analysis in \cref{sec:CT_QE}. At almost all values of $a_u$ for all indices, \cref{tab:p_LR_ci} shows no significant difference between the goodness of fit of the common intensity model compared with the bivariate model. This greatly expands upon the same finding in \cite{Tomlinson2021}, which only tested the SPX at $a_u=0.025$. This suggest that a common intensity arrivals process can be universally assumed for the extreme daily log-returns of large cap stock indices. \cref{tab:p_LR_c} shows that the $a_{\lambda}^{\rsTA} = 2 a_u \rsUdt^{-1}$ constraint produces a negligible cost to the goodness of fit in all but \ENone{} case out of 120 in the in-sample data. \cref{tab:p_LR_t} shows that the Student-$t$ distributed bulk achieves a significantly better fit than the normal distribution in most cases. Given these results, we use the common intensity exceedance model with the $a_{\lambda}^{\rsTA} = 2 a_u \rsUdt^{-1}$ constraint applied and with a Student-$t$ distributed bulk in \cref{sec:Data,sec:CT_QE}.

\begin{table*}
\caption{\label{tab:p_LR_ci}
Likelihood ratio test $p$-values comparing the goodness of fit to threshold exceeding log-returns of the common-intensity model $H_{2}{\left(a_u\right)}$ versus the bivariate model $H_{2,\mathrm{bi}}{\left(a_u\right)}$. $\mathcal{H}_0: \ell_{u}{\left[H_{2}{\left(a_u \right)}\right]} = \ell_{u}{\left[H_{2,\mathrm{bi}}{\left(a_u\right)}\right]}$. $\mathcal{H}_1: \ell_{u}{\left[H_{2}{\left(a_u\right)}\right]} < \ell_{u}{\left[H_{2,\mathrm{bi}}{\left(a_u\right)}\right]}$. Rejections of $\mathcal{H}_0$ at the 95\% and 99\% confidence levels are highlighted in light and dark gr\ENe{}y, respectively.}
\FTABpLRci
\end{table*}
\begin{table*}
\caption{\label{tab:p_LR_c}
Likelihood ratio test $p$-values comparing the goodness of fit to threshold exceeding returns of $H_{2}{\left(a_u\right)}$ with and without the constraint $a_{\lambda}^{\rsTA} = 2 a_u \rsUdt^{-1}$. $\mathcal{H}_0: \ell_{u}{\left[H_{2}{\left(a_u \middle| a_{\lambda}^{\rsTA} = 2 a_u \rsUdt^{-1}\right)}\right]} = \ell_{u}{\left[H_{2}{\left(a_u \middle| a_{\lambda}^{\rsTA} \neq 2 a_u \rsUdt^{-1}\right)}\right]}$. $\mathcal{H}_1: \ell_{u}{\left[H_{2}{\left(a_u \middle| a_{\lambda}^{\rsTA} = 2 a_u \rsUdt^{-1}\right)}\right]} < \ell_{u}{\left[H_{2}{\left(a_u \middle| a_{\lambda}^{\rsTA} \neq 2 a_u \rsUdt^{-1}\right)}\right]}$. Rejections of $\mathcal{H}_0$ at the 95\% and 99\% confidence levels are highlighted in light and dark gr\ENe{}y, respectively.}
\FTABpLRc
\end{table*}

\begin{table*}
\caption{\label{tab:p_LR_t}
Likelihood ratio test $p$-values comparing the goodness of fit to bulk returns of $H_{2}^{\rsDistNorm}{\left(a_u\right)}$ against $H_{2}^{\rsDistT}{\left(a_u\right)}$. $\mathcal{H}_0: \ell_{B}{\left[H_{2}^{\rsDistNorm}{\left(a_u\right)}\right]} = \ell_{B}{\left[H_{2}^{\rsDistT}{\left(a_u\right)}\right]}$. $\mathcal{H}_1: \ell_{B}{\left[H_{2}^{\rsDistNorm}{\left(a_u\right)}\right]} < \ell_{B}{\left[H_{2}^{\rsDistT}{\left(a_u\right)}\right]}$. Rejections of $\mathcal{H}_0$ at the 95\% and 99\% confidence levels are highlighted in light and dark gr\ENe{}y, respectively.}
\FTABpLRt
\end{table*}

\BIBprint

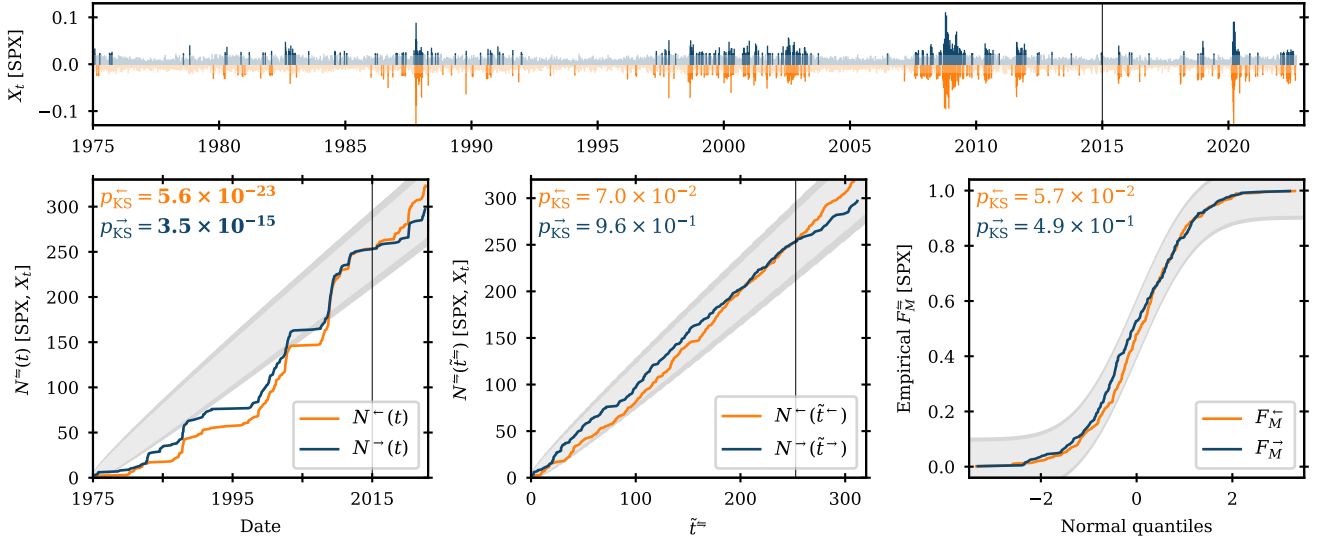
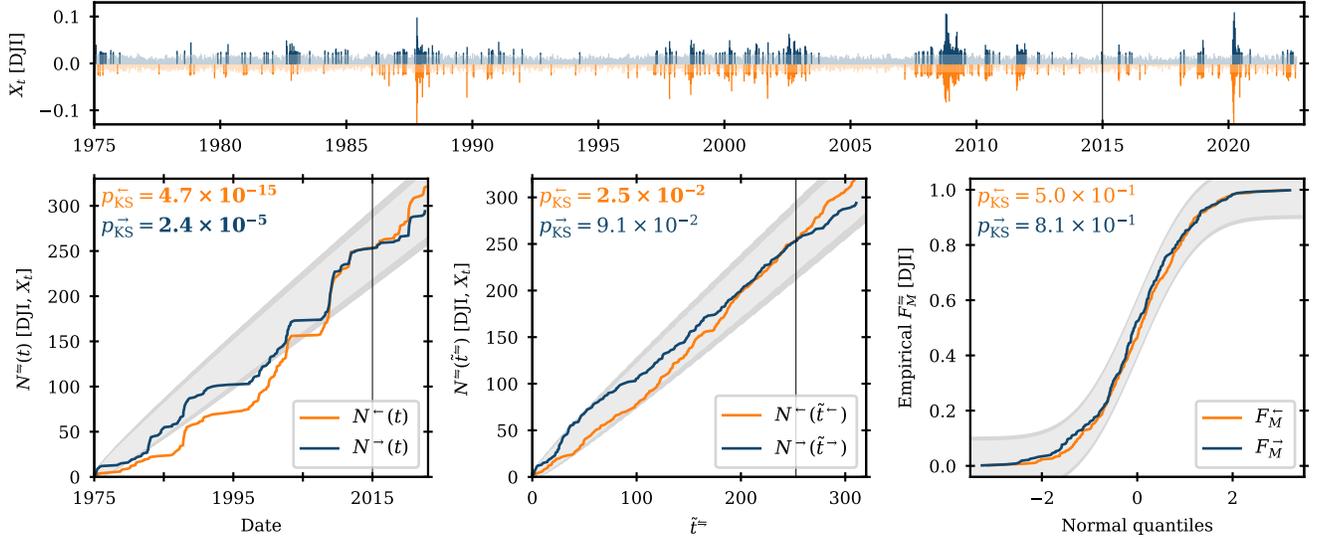
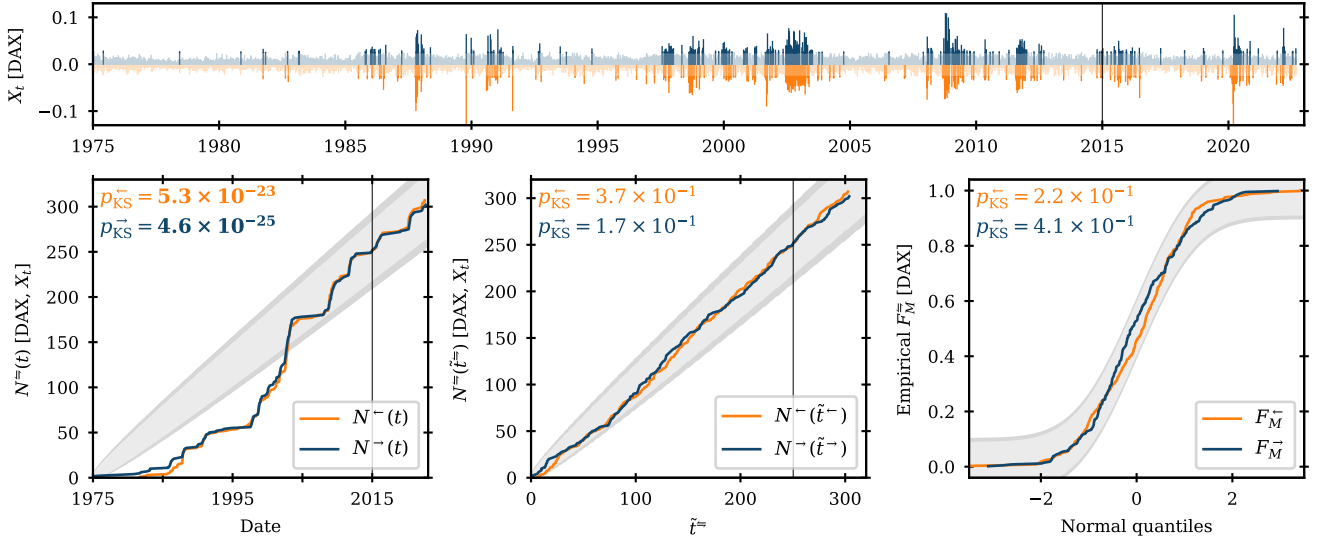
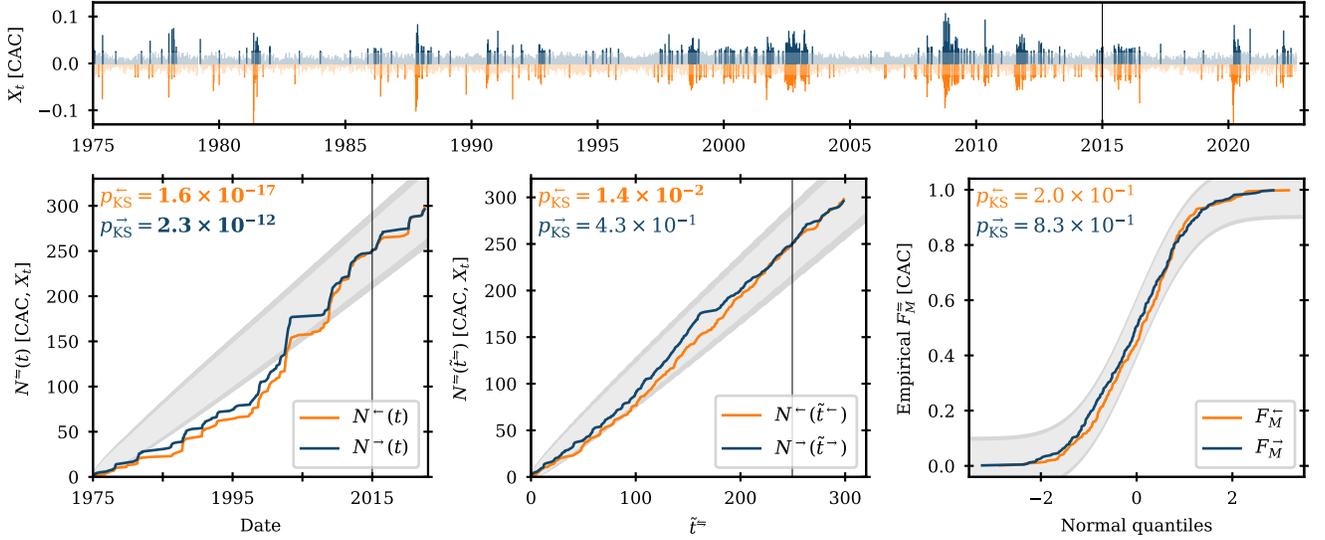
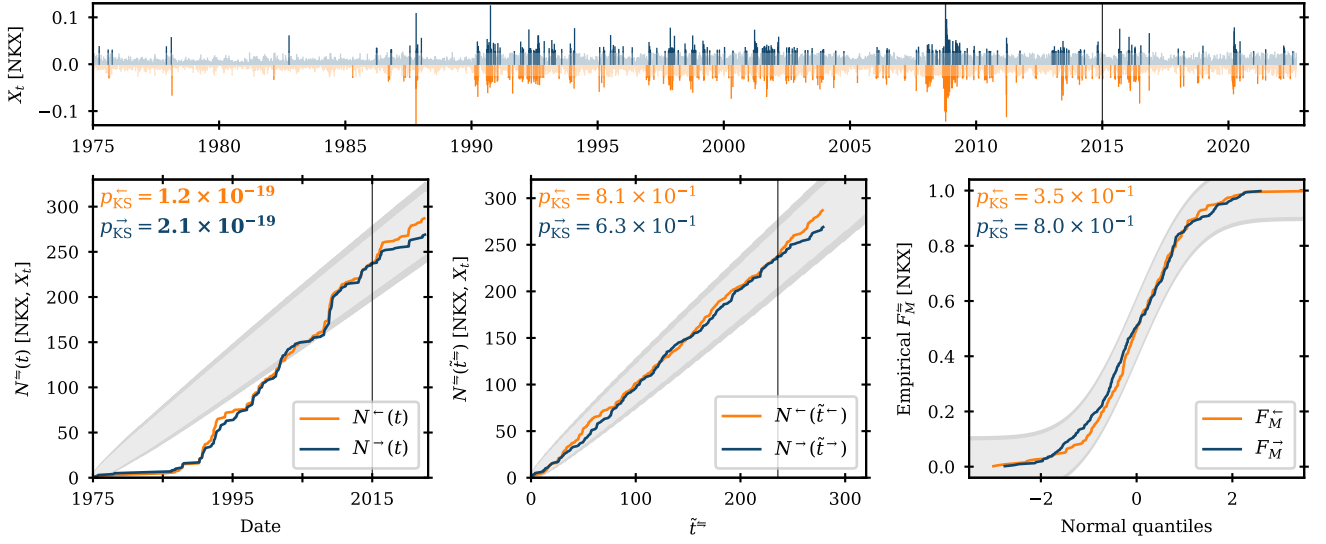
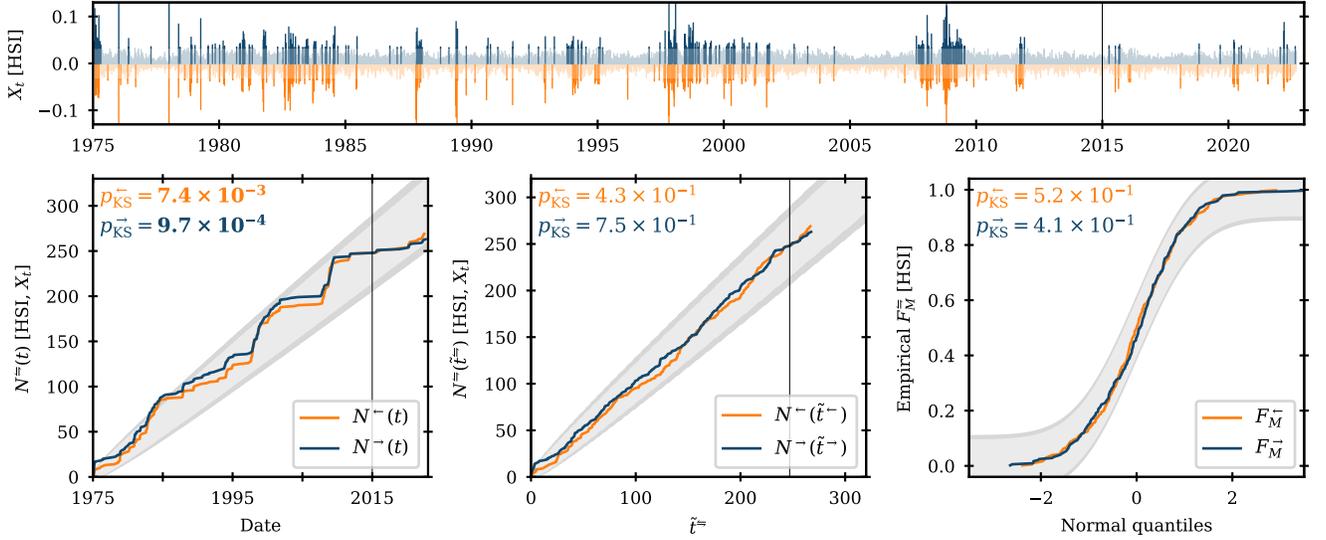
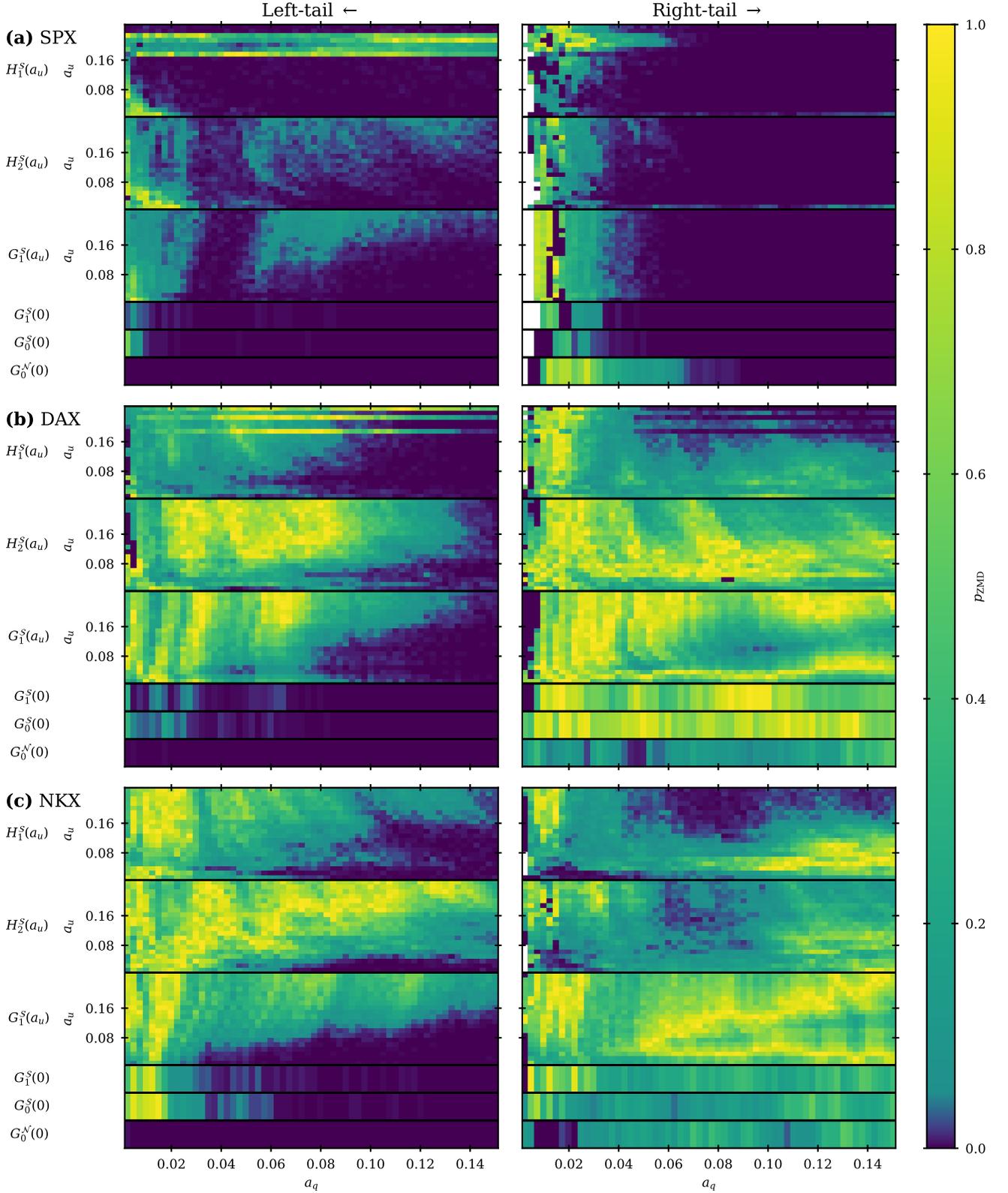
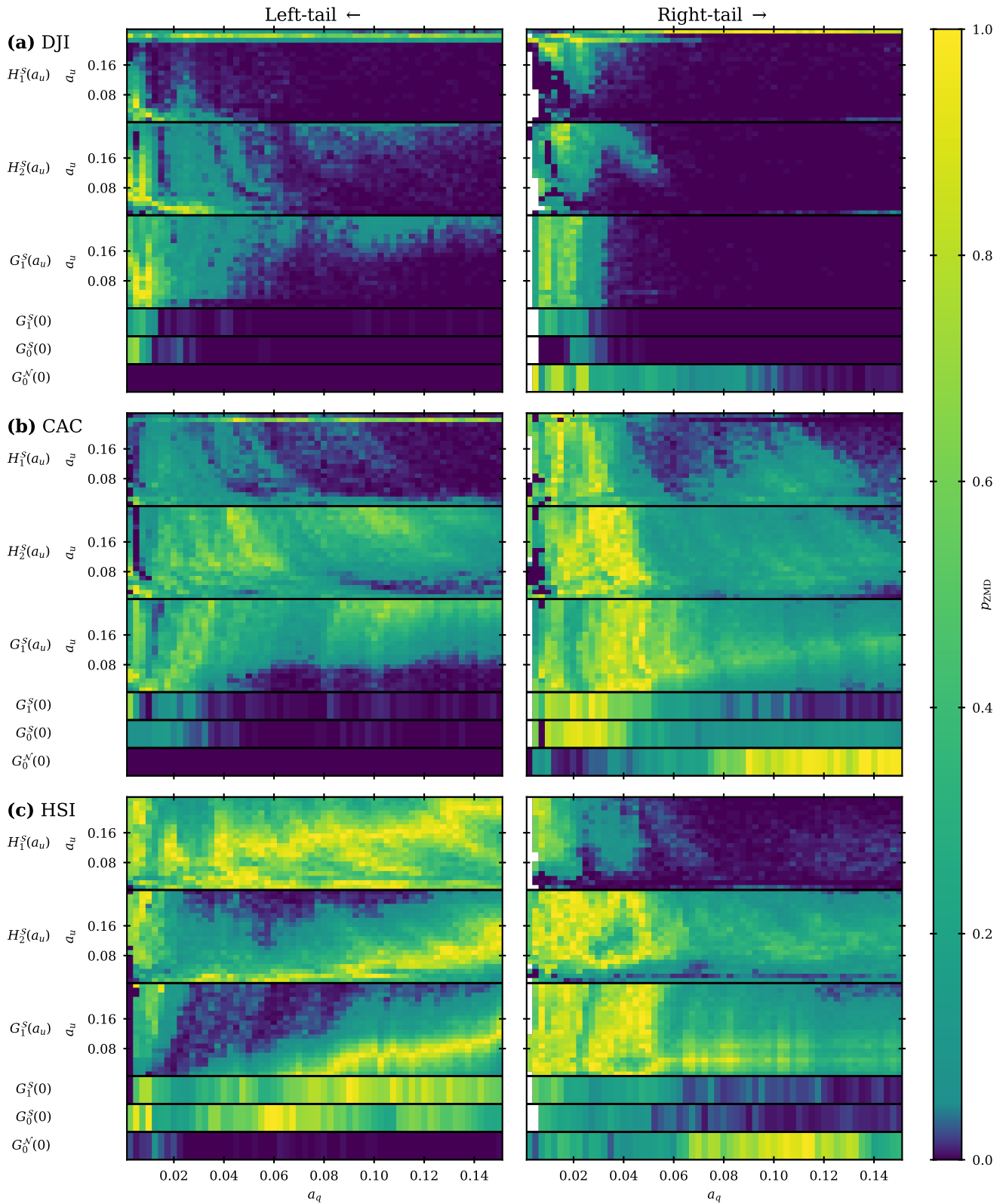
\end{document}